\documentclass[11pt]{article}
\usepackage[utf8]{inputenc}
\usepackage[T1]{fontenc}
\usepackage{booktabs}
\usepackage[numbers,sort&compress]{natbib}

\usepackage{authblk}

\newcommand{\revOne}[1]{\normalcolor #1}
\newcommand{\revTwo}[1]{\normalcolor #1}
\newcommand{\revBoth}[1]{\normalcolor #1}
\newcommand{\revSelf}[1]{\normalcolor #1}

\usepackage{amsmath,amsfonts}
\usepackage{physics}
\usepackage{orcidlink}
\usepackage{qcircuit}
\usepackage{subcaption}

\usepackage{cellspace} 
\newtheorem{theorem}{Theorem}[section]
\newtheorem{lemma}[theorem]{Lemma}

\title{Stabilizer Code-Generic Universal Fault-Tolerant Quantum Computation}

\author[1,*]{Nicholas J.C. Papadopoulos\orcidlink{0000-0002-6357-0030}}
\author[1]{Ramin Ayanzadeh\orcidlink{0000-0001-6687-5668}}
\affil{University of Colorado Boulder, Department of Computer Science, Boulder, CO, 80309, USA}
\affil[*]{Corresponding author. Email: nicholas.papadopoulos@colorado.edu}
\date{}

\begin{document}

\flushbottom
\maketitle

\thispagestyle{empty}

\begin{abstract}
Fault-tolerant quantum computation allows quantum computations to be carried out while resisting unwanted noise.
Several error-correcting codes have been developed to achieve this task, but none alone are capable of \revSelf{transversal} universal quantum computation.
This universality is highly desired and often achieved using techniques such as code concatenation, code switching, magic state distillation, or pieceable fault tolerance. 
\revSelf{Existing constructions can incur substantial overhead or rely on properties specific to particular codes.
This work proposes implementing logical Clifford and T gates through novel, ancilla-mediated, pieceable interactions to construct a universal fault-tolerant quantum gate set.
Our construction is deterministic, does not modify the underlying data codes or registers, and is generic over all \revOne{data} stabilizer codes.}
Thus, any single code becomes capable of universal quantum computation by leveraging helper codes in ancilla registers and mid-circuit measurements.
Furthermore, since these logical gates are stabilizer code-generic, these implementations enable communication between heterogeneous stabilizer codes.
\end{abstract}

\section{Introduction}

Quantum computing promises exponential speedups for certain classes of problems, but its practical realization remains unreliable due to the fragile nature of quantum information.
Quantum bits (qubits) are susceptible to decoherence and operational noise, necessitating error-resistant protocols.
Quantum error correction (QEC) uses redundant information to protect against such error-causing noise by encoding the information of a quantum state using a collection of qubits, called a logical qubit~\cite{discreteErrorsFromContinuous,gottesmanStabilizers,gottesmanStabilizerFormalism,repetitionCode,hammingCode,shorsCode,generalizedShorCode,baconShor,surfaceCode,reedMuller,colorCode,five-qubit,four-qubit,dodecacode}.
Fault-tolerant protocols use these encodings to perform error-resistant quantum computations~\cite{shorFtComputation,shorFtGeneralization,nielsen_chuang_2010_fault_tolerance,colorCodesFt,subsystemFtBaconShor}.
A central challenge in fault-tolerant quantum computing is efficiently implementing a set of gates that can perform any quantum computation, called a universal gate set~\cite{uniFtNoKnots,uniFtQcTransversalQec,uniFtQcWithStabilizers,uniFtQcWithBaconShor,colorCodeUniGates,butt2024measurement}.

All QEC codes support only a limited set of transversal logical gates.
Logical gates have the effect of a physical gate on logical qubits, and transversal means that they can be performed bit-wise with physical gates and are inherently fault-tolerant.
Hence, additional techniques must be used to fault-tolerantly perform any non-native logical gates.
Established techniques often include code concatenation~\cite{concatenatedCodes}, code switching~\cite{codeSwitching}, magic state distillation~\cite{magicStates}, \revSelf{and} pieceable fault tolerance~\cite{pieceableFt}.
However, these techniques are often special procedures only available on the codes for which they were designed, they remain experimentally demanding, or they otherwise have heavy restrictions associated with them.

\revSelf{Code concatenation nests codes within one another to build arbitrary distance, but achieving universal FTQC this way requires structural compatibility between levels and high overheads when scaling~\cite{concatenatedCodes}.
Code switching~\cite{codeSwitching,conversionSteaneReedMuller,codeSwitchingNearTerm,codeSwitchingQldpcDimensions,heterogeneousByBus,ouyang2024measurement} transfers information between differently encoded registers to utilize specific properties from each but is often limited to specific codes that work with a particular procedure.}
Magic state distillation~\cite{magicStateOriginal,magicStates} prepares many noisy copies of special quantum states that can then be used to output fewer copies of the quantum state with higher fidelity.
These purified quantum states can then implement non-Clifford gates via gate teleportation~\cite{teleportation}.
Although recent improvements to this process have made it far more efficient than previously~\cite{msdLessThanCnot,magicStateNotAsCostly,constantOverheadMagicStateDistillation,qldpcMagicStateDistillation}, it is still often considered costly; Stein et al. describe it as having ``expensive, if not prohibitive cost when scaling''~\cite{heterogeneousByBus}.
Standard magic state distillation is nondeterministic \revSelf{due to post-selection}, making the process unpredictable and costly.
\revSelf{While deterministic variants of magic state distillation exist, recent approaches, such as that of Heußen~\cite{heussen2025magic}, involve certain trade-offs, particularly regarding reduced noise suppression rate per round.}
\revSelf{While this work constitutes a form of pieceable fault tolerance, previous constructions of pieceable fault tolerance often relied on decomposing gates into fault-tolerant segments tailored to the target code.}
Many of the guarantees and algorithms given for pieceable fault tolerance are restricted to nondegenerate codes that already possess a complete set of fault-tolerant Clifford gates.

To address these limitations, we propose a novel, stabilizer code-generic (SCG) framework that achieves universality via a strategy we term \textit{ancilla mediation}.
\revOne{We define ancilla mediation as a technique in which auxiliary encoded blocks mediate logical transformations on data blocks, while the logical data remain in their original registers and the stabilizer group of each data code remains unchanged throughout the procedure.}
\revSelf{We achieve universality despite the restriction that no single code can have a transversal implementation of a universal gate set~\cite{eastinKnill} by decoupling the universal-gate implementation from the specific structure of the data code.}
In contrast with code concatenation and code switching, we achieve universal fault-tolerant quantum computation on arbitrary, heterogeneous stabilizer codes (i.e., stabilizer code-generic).
The data information always remains in its initial codes and registers, thereby preserving the properties of the underlying codes, including their distances and error-correcting capabilities.
In contrast with standard, post-selected magic state distillation, our technique is deterministic\revBoth{, and creating a $T$ state for teleportation can be done with a single, transversal step}.
In contrast to pieceable fault tolerance, the SCG circuits shown in this work \revOne{change only by scaling the distance and using the correct logical $\bar{X}/\bar{Z}$ gates of the data codes.
Furthermore, the SCG gates have no restriction on the type of stabilizer code on which they act.}

We target the Clifford+$T$ logical gate set, which is a known universal gate set~\cite{cliffordPlusNonclifford,latticeSurgery,magicStateRobustness}, while also supporting operation between data encoded in different QEC codes.
Both of these tasks are achieved through the use of ancilla registers encoded with the generalized Shor code ($GSC$)~\cite{generalizedShorCode}.
The particularly useful properties of this code enable transversal logical controlled-X and controlled-Z gates targeting any stabilizer code.
In addition to $GSC$, one additional stabilizer code that has a fault-tolerant implementation of a $T$ logical gate must be used to achieve the full, universal set.
For example, triorthogonal codes are a well-documented class of such codes~\cite{triorthogonal,triorthogonal_generation}.
\revOne{Although the helper codes must have specific properties, the only required property of the data codes is that they are stabilizer codes.
We define stabilizer code-generic universal quantum computation as the ability to perform universal quantum computation on $k$ logical qubits $\mathcal{Q}_1,\mathcal{Q}_2,\dots,\mathcal{Q}_k$, where each logical qubit may be encoded in an arbitrary stabilizer code, with up to $k$ distinct data-code encodings.
Other, non-generic codes can be used outside of the data codes.
}

\revOne{Following standard definitions of fault tolerance~\cite{nielsen_chuang_2010_fault_tolerance,gottesmanQecIntro,rosenblum2018fault,delfosse2022beyond}, we call a procedure fault-tolerant if a single fault at any circuit location results in at most one error in each encoded block output by the procedure.}
\revSelf{Throughout this work, we assume fault-tolerant preparation of encoded states and fault-tolerant measurement schemes.
These standard fault-tolerant primitives are treated as available for the constituent codes, while our analysis focuses on the additional fault-tolerance requirements introduced by the SCG interactions.}

We first explain how to use the \revBoth{X-basis} of $GSC$, to which we refer as $\revBoth{GSCX}$, to fault-tolerantly control logical $\bar{X}/\bar{Z}$ gates.
A control register encoded in $\revBoth{GSCX}$ can target an arbitrary data code\revOne{, given the data code's logical $\bar{X}/\bar{Z}$ gates}.
This extends to allow an SCG Hadamard gate on any data encoding, as this gate can be composed of only logical controlled-$\bar{X}/\bar{Z}$ gates controlled by an ancilla register.
We then describe how to construct an SCG controlled-$\bar{X}/\bar{Z}$ gate between any two data encodings using the same techniques as are used in the SCG Hadamard gate.
Finally, we show that an SCG rotation about the Z-axis, which includes an SCG $\revSelf{\bar{T}}$ gate, can be performed by entangling the data encoding, using \revSelf{other SCG gates}, with a code that can perform the desired Z-rotation fault-tolerantly.
Constructing these logical gates, therefore, satisfies the Clifford+$T$ universal gate set.
\revSelf{For the SCG $\bar{H}$ and controlled-$\bar{X}/\bar{Z}$} gates, we first describe a fault-intolerant version as a blueprint for the gates we want to achieve, then we use ancilla mediation with $\revBoth{GSCX}$ to perform the blueprint fault-tolerantly.

By decoupling logical gates from code-specific constraints, this framework, using exclusively ancilla-mediated protocols, (1) enables universal fault-tolerant quantum computing for any stabilizer code and (2) supports diverse QEC codes in heterogeneous systems, where distinct \revOne{data encodings can communicate without requiring modification or adaptation of the data codes themselves.}
The generic nature of this framework allows its incorporation into any stabilizer code and eliminates the need to derive specific universal gate implementations.
\revOne{Hence, the framework can be applied in future studies of existing and yet-undiscovered QEC codes and their combinations.
Furthermore, by enabling communication between arbitrary stabilizer data codes without modifying their underlying encodings, the framework could support distributed quantum computing in which different hardware platforms employ different QEC codes.}

\section{Results}

\subsection{Notation}

Table~\ref{tab:notation-states} summarizes the following notation used throughout this work. 
We denote the $i$th logical qubit as $\mathcal{Q}_i$.
This notation allows for heterogeneity; $\mathcal{Q}_1$ and $\mathcal{Q}_2$ may use different codes or architectures.
For instance, $\mathcal{Q}_1$ could be in a surface code while $\mathcal{Q}_2$ is in a Steane code,
or $\mathcal{Q}_1$ and $\mathcal{Q}_2$ could be the first and second qubit, respectively, of a four-qubit code ($[[4, 2, 2]]$)~\cite{four-qubit}.

Let $\mathbb{Z}$ denote the set of integers and $\mathbb{C}$ denote the set of complex numbers.
Also let $\ket{\psi}_{\mathcal{Q}_i} = \alpha \ket{0}_{\mathcal{Q}_i} + \beta \ket{1}_{\mathcal{Q}_i}$, for some $\alpha, \beta \in \mathbb{C}$ and $|\alpha|^2 + |\beta|^2 = 1$, be the state of some encoded logical qubit.
Similarly, let $\ket{\phi}_{\mathcal{Q}_i} = \alpha' \ket{0}_{\mathcal{Q}_i} + \beta' \ket{1}_{\mathcal{Q}_i}$ be the state of some different logical qubit.
The subscript of a ket denotes the register of that state, encoded with a stabilizer code, which may be $\mathcal{Q}_i$ for some arbitrary code, or it may be specified with the abbreviation of a particular code.

Let \revSelf{$U$} denote an arbitrary, single-qubit gate so that $\bar{\revSelf{U}}_{\mathcal{Q}_i}$ denotes its respective logical gate acting on logical qubit $\mathcal{Q}_i$.
$\overline{C\revSelf{U}}_{\mathcal{Q}_1,\mathcal{Q}_2}$ represents a logical controlled gate, controlled by $\mathcal{Q}_1$, targeting $\mathcal{Q}_2$.

\begin{table}
  \centering
  \caption{Notation used throughout the paper.}
  \label{tab:notation-states}
  \begin{tabular}{Sl Sc}
    \toprule
    \textbf{Notation} & \textbf{Meaning} \\
    \midrule
    $\mathcal{Q}_i$ & \begin{tabular}{c}
        A logical qubit \\
        encoded with an arbitrary stabilizer code
    \end{tabular} \\
    \hline
    $\ket{\psi}_{\mathcal{Q}_i}$ & $\alpha \ket{0}_{\mathcal{Q}_i} + \beta \ket{1}_{\mathcal{Q}_i}$ \\
    \hline
    $U$ & An arbitrary gate \\
    \hline
    $\bar{U}_{\mathcal{Q}_i}$ & Logical gate $U$ acting on $\mathcal{Q}_i$ \\
    \hline
    $\overline{CU}_{\mathcal{Q}_i, \mathcal{Q}_j}$ & \begin{tabular}{c}
         Logical gate $U$ \\
         acting on $\mathcal{Q}_j$ and controlled by $\mathcal{Q}_i$
    \end{tabular} \\
  \bottomrule
\end{tabular}
\end{table}

\subsection{The Generalized Shor Code and its \revBoth{X-Basis}}

\subsubsection{GSC}

\begin{figure}[t!]
    \centering
    \includegraphics[width=.55\linewidth]{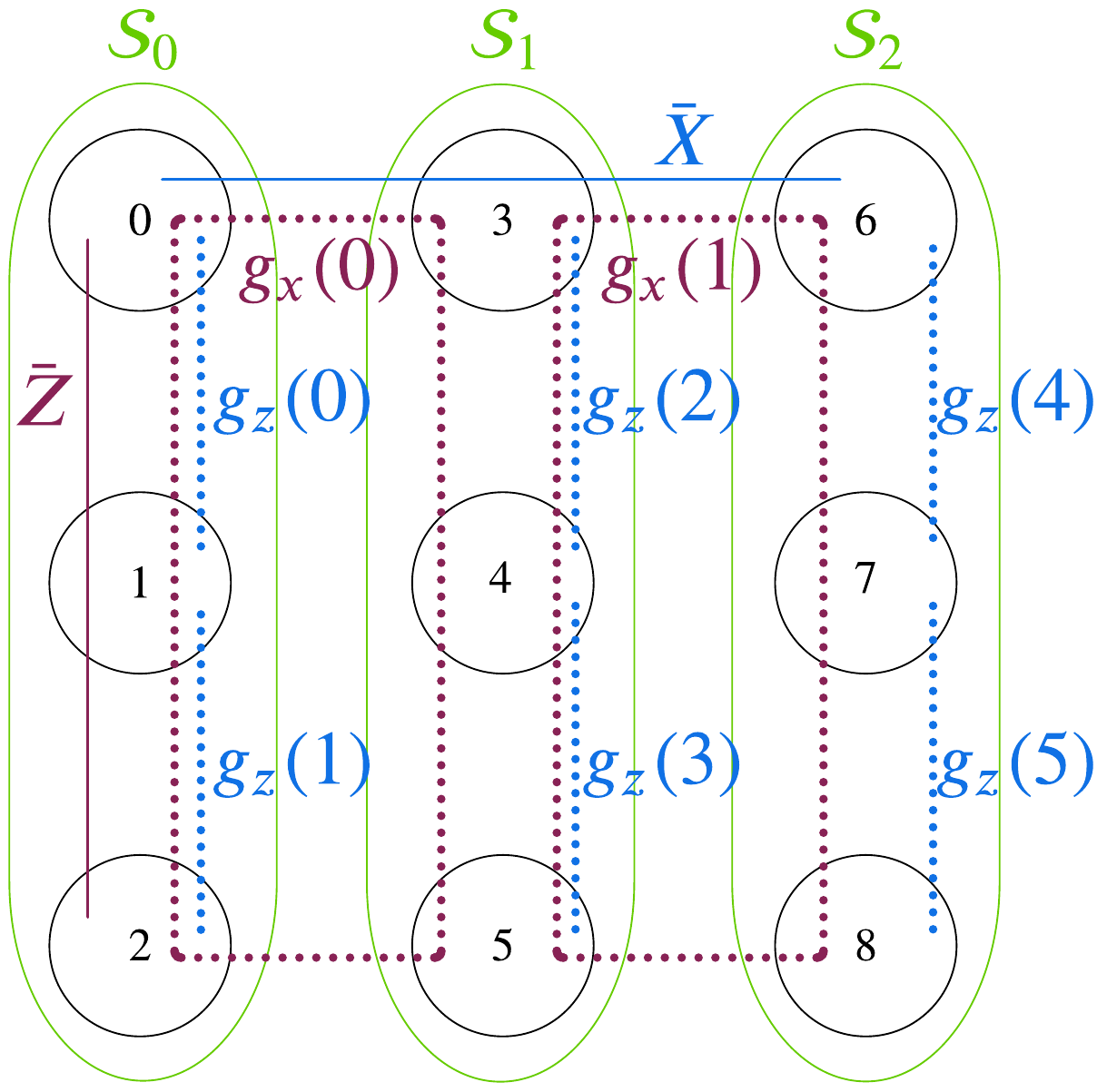}
    \caption{$GSC_{3,3}$ in a grid structure.
    Data qubits are represented by black circles with center labels.
    The $i$th subregister, $\mathcal{S}_i$, is labeled and circled in green.
    Red operators indicate a collection of X gates, while blue operators indicate a collection of Z gates.
    Hence, the qubits used for the $\bar{X}$ logical gate are shown intersecting the horizontal, solid blue line, and those used for the $\bar{Z}$ logical gate are shown intersecting the vertical, solid red line.
    Those used for the X-stabilizers are shown intersecting the dotted, red squares, and those used for the Z-stabilizers are shown intersecting the dotted, blue lines.}
    \label{fig:mcc-example}
\end{figure}

$GSC$ generalizes the Shor code to have any number of ``cat'' states $\frac{1}{\sqrt{2}}(\ket{0 \dots 0} + \ket{1 \dots 1})$ (named after Schrödinger’s cat, also known as a GHZ state~\cite{ghzState}) along with any number of qubits per cat state.
Cat states can be encoded using any number of initialization techniques~\cite{catCreation}.
Let $GSC$ take two parameters: the number of cat states, $a$, and the number of qubits per cat state, $b$.
We mark these in subscripts when referring to a specific $GSC$, $GSC_{a,b}$\revSelf{, and call each of the cat states a subregister of the code.} 
$\mathcal{S}_i$ denotes the $i$th subregister.
The logical computational basis states $\ket{x}_{GSC_{a,b}}$, for $x \in \{0,1\}$ specifying the logical state $\ket{x}$ encoded with $GSC_{a,b}$, are therefore
\begin{equation}
    \begin{aligned}
        \ket{x}_{GSC_{a,b}} &= \frac{1}{\sqrt{2^{a}}} \left( \ket{0}^{\otimes b} + (-1)^x \ket{1}^{\otimes b} \right)^{\otimes a}.
    \end{aligned}
\end{equation}
For example, the original Shor code is a $GSC_{3,3}$ code with computational basis states
\begin{equation}
    \begin{aligned}
        \ket{0}_{GSC_{3,3}} &= \frac{(\ket{000} + \ket{111}) (\ket{000} + \ket{111}) (\ket{000} + \ket{111})}{2\sqrt{2}} \\
        \ket{1}_{GSC_{3,3}} &= \frac{(\ket{000} - \ket{111}) (\ket{000} - \ket{111}) (\ket{000} - \ket{111})}{2\sqrt{2}}. \\
    \end{aligned}
\end{equation}
$GSC$ has Z-stabilizers consisting of Z gates on pairs of adjacent qubits in each subregister, formally given by Eq.~\eqref{eq:multi-cat-stabilizers-z} with $0 \le i < a(b - 1)$,
\begin{equation}\label{eq:multi-cat-stabilizers-z}
    \begin{aligned}
        g_z(i, b) &= Z_{q(i, b)} Z_{q(i, b) + 1}
    \end{aligned}
\end{equation}
where $q(i, b) = \lfloor i / (b - 1) \rfloor b + (i \mod{(b - 1}))$.
X-stabilizers consist of X gates on all qubits in pairs of adjacent subregisters, formally given by Eq.~\eqref{eq:multi-cat-stabilizers-x} with $0 \le i < a - 1$,
\begin{equation}\label{eq:multi-cat-stabilizers-x}
    \begin{aligned}
        g_x(i) &= \prod_{j = bi}^{bi + 2b - 1} X_j.
    \end{aligned}
\end{equation}
In equations~\eqref{eq:multi-cat-stabilizers-z} and~\eqref{eq:multi-cat-stabilizers-x}, $Z_j$ and $X_j$ are $Z$ and $X$ single-qubit gates acting on the $j$th qubit of the register.
Note that while $GSC_{a,b}$ shares the physical layout of the Bacon-Shor subsystem code~\cite{baconShor}, it is defined here as a pure stabilizer code.
Specifically, $GSC_{a,b}$ represents a fixed-gauge limit of the Bacon-Shor framework, where the gauge degrees of freedom are fully constrained by the stabilizer group.
The operators for $GSC_{a,b}$ are
\begin{equation}
    \begin{aligned}
        \bar{Z} &= \prod_{j = 0}^{b - 1} X_j,
        \quad \bar{X} = \prod_{j = 0}^{a - 1} Z_{\revSelf{b}j}.
    \end{aligned}
\end{equation}
An example of these stabilizers can be found in Supplementary Note 1.1,
and an image depicting $GSC_{3,3}$ is shown in Fig.~\ref{fig:mcc-example}.

\subsubsection{\revBoth{The X-Basis of GSC}}
\revBoth{We choose to think of the X-basis of $GSC$, labeled $GSCX$, as the primary basis convention because letting the computational basis represent $GSCX$'s Z-basis simplifies many of the expressions and circuits.
$\revBoth{GSCX}$, similarly to $GSC$, takes $a$ and $b$ as parameters and can be labeled $\revBoth{GSCX}_{a,b}$.
Since $\revBoth{GSCX}$ serves as $GSC$'s X-basis rather than as a distinct code, the two have the same stabilizers and codespace while the roles of their logical $\bar{X}$ and $\bar{Z}$ operators are switched.}
\begin{equation}
    \begin{aligned}
        \bar{Z}_{\revBoth{GSCX}_{a,b}} &= \bar{X}_{GSC{a,b}}, \quad \bar{X}_{\revBoth{GSCX}_{a,b}} = \bar{Z}_{GSC_{a,b}}.
    \end{aligned}
\end{equation}
The logical computational basis states $\ket{x}_{\revBoth{GSCX}_{a,b}}$, for $x \in \{0,1\}$, are
\begin{equation}
    \begin{aligned}
        \ket{x}_{\revBoth{GSCX}_{a,b}} &= \revSelf{\frac{1}{\sqrt{2^{a-1}}}} \sum_{\substack{0 \leq i < 2^a \\ \mathrm{wt}(i) \equiv x \pmod{2}}} \bigotimes_{j=0}^{a-1} \ket{i_j}^{\otimes b}
    \end{aligned}
\end{equation}
where $\mathrm{wt}(i)$ denotes the Hamming weight of $i$, i.e., the number of 1s in its binary representation, and $i_j = \left\lfloor \frac{i}{2^j} \right\rfloor \bmod 2$, i.e., the $j$th bit of $i$.
Essentially, $\ket{0}_{\revBoth{GSCX}_{a,b}}$ consists of the basis states in $\ket{0}_{GSC_{a,b}}$ that have an even number of subregisters in the $\ket{1 \dots 1}$ state, and $\ket{1}_{\revBoth{GSCX}_{a,b}}$ consists of the remaining basis states in $\ket{0}_{GSC_{a,b}}$.
An example of these states can be more explicitly seen in Supplementary Note 1.1.

\subsection{Controlled-X/Z Using \revBoth{GSCX}}\label{sec:cx-by-cpc}

Throughout the rest of this paper, we imply the parameters $a$ and $b$ from context when $GSC$ and $\revBoth{GSCX}$ appear without subscripts.
Unless stated otherwise, let $a,b \in \mathbb{Z} \geq 3$, and let $b \geq \max\left( \left| \bar{X}_{\mathcal{Q}_1} \right|, \left| \bar{Z}_{\mathcal{Q}_1} \right| \right)$.
Here, $|\bar{\revSelf{U}}_{\mathcal{Q}_i}|$ denotes the number of physical non-identity gates comprising $\bar{\revSelf{U}}_{\mathcal{Q}_1}$.
These constraints allow for correction of at least one arbitrary error and enough qubits per cat state to act transversally on the target.

$\revBoth{GSCX}$ has a useful capability to act as a control register targeting any data encoding when performing logical controlled-$\bar{X}/\bar{Z}$ gates.
Similarly to Shor's measurement procedure~\cite{nielsen_chuang_2010_ft_measurement,numMeasurementRoundsForShorStyle}, it uses its subregisters to perform the data code's logical operator in a bit-wise fashion.

\begin{lemma}\label{lem:cx}
    Given the initial state $\ket{\psi}_{\revBoth{GSCX}} \ket{\phi}_{\mathcal{Q}_1}$, \revOne{the circuit shown in Fig.~\ref{fig:gsch-sequence} fault-tolerantly performs the logical gate $\overline{CO}_{\revBoth{GSCX}, \mathcal{Q}_1}$ for any gate $O \in \{ X, Z \}$, such that
    \begin{equation}
        \begin{aligned}
            \overline{CO}_{\revBoth{GSCX}, \mathcal{Q}_1}\ket{\psi}_{\revBoth{GSCX}} \ket{\phi}_{\mathcal{Q}_1} &= \alpha \ket{0}_{\revBoth{GSCX}} \ket{\phi}_{\mathcal{Q}_1} + \beta \ket{1}_{\revBoth{GSCX}} \left( \bar{O}_{\mathcal{Q}_1} \ket{\phi}_{\mathcal{Q}_1} \right).
        \end{aligned}
    \end{equation}}
\end{lemma}

\revOne{In the same vein as Shor's measurement procedure, each subregister $\mathcal{S}_i$ of $\revBoth{GSCX}$ can transversally perform $\overline{CO}_{\mathcal{S}_i,\mathcal{Q}_1}$, as shown in Fig.~\ref{fig:transversal-detail}.
Applying these interactions sequentially over all subregisters gives the full logical operation shown in Fig.~\ref{fig:gsch-sequence}.}
\revSelf{$\ket{0}_{\revBoth{GSCX}}$ has an even number of subregisters in the $\ket{1 \dots 1}$ state, while $\ket{1}_{\revBoth{GSCX}}$ has an odd number.
For each subregister in the $\ket{1 \dots 1}$ state, $\bar{O}_{\mathcal{Q}_1}$ is applied once, and $\left(\bar{O}_{\mathcal{Q}_1}\right)^2 = \mathbb{I}$.
Thus, the resulting action depends only on the parity of these subregisters,
so that
\begin{equation}
    \prod^{a-1}_{j=0} \overline{CO}_{\mathcal{S}_j, \mathcal{Q}_1} \ket{x}_{\revBoth{GSCX}} \ket{\phi}_{\mathcal{Q}_1} = \ket{x}_{\revBoth{GSCX}} \left( \left(\bar{O}_{\mathcal{Q}_1}\right)^x \ket{\phi}_{\mathcal{Q}_1} \right),
\end{equation}
where $x \in \{0,1\}$.
}

Error correction done between each logical gate maintains the state.
However, in order to perform the intermediate error corrections, the X-stabilizers of $\revBoth{GSCX}_{a,b}$ need to be temporarily modified after each $i < a - 1$ since the codespace is temporarily changed until all iterations are performed.
Z-stabilizers require no modification because they are contained within individual subregisters, whereas the modification of the codespace is grouped at the subregister level.
The modified $j$th X-stabilizer after each $i$ is (see Supplementary Note 2.1)
\begin{equation}\label{eq:modified-stabilizers}
    \begin{aligned}
        g_x'(i, j) = g_x(j) \left( \bar{O}_{\mathcal{Q}_1} \right)^{\delta_{ij}},
    \end{aligned}
\end{equation}
where $\delta_{ij} = \begin{cases}1 & i = j \\ 0 & i \ne j\end{cases}$ is the Kronecker delta function.
After the last $\overline{CO}_{\mathcal{S}_i, \mathcal{Q}_1}$, the stabilizers return to normal.
\begin{figure}
    \centering
    \captionsetup{justification=raggedright,singlelinecheck=false}
    
    \begin{subfigure}{\textwidth}
        \centering
        \caption{\revOne{Single-subregister transversal interaction.}}
        \label{fig:transversal-detail}

        \centerline{
            \Qcircuit @C=1.5em @R=1em {
                \lstick{d_{0}} & \gate{O_0} & \qw      & \qw      & \qw      & \qw \\
                \lstick{d_{1}} & \qw      & \gate{O_1} & \qw      & \qw      & \qw \\
                \lstick{\vdots} & & & \ddots & & \\
                \lstick{d_{b - 1}} & \qw      & \qw      & \qw      & \gate{O_{b - 1}} & \qw \\
                \lstick{s_{0}} & \ctrl{-4} & \qw      & \qw      & \qw      & \qw \\
                \lstick{s_{1}} & \qw      & \ctrl{-4} & \qw      & \qw      & \qw \\
                \lstick{\vdots} & & & \ddots & & \\
                \lstick{s_{b - 1}} & \qw      & \qw      & \qw      & \ctrl{-4} & \qw 
                %
                \inputgroupv{2}{3}{5.5em}{1em}
                    {\ket{\phi}_{\mathcal{Q}_1} \hspace{5em}}
                \inputgroupv{7}{7}{5.5em}{0.2em}
                    {\mathcal{S}_i \hspace{5em}}
            }
        }
    \end{subfigure}

    \vspace{2em}

    \begin{subfigure}{\textwidth}
        \centering
        \caption{\revOne{Full logical interaction across all $\revBoth{GSCX}$ subregisters.}}
        \label{fig:gsch-sequence}
        \makebox[\textwidth][r]{
            \Qcircuit @C=0.8em @R=1.2em {
                \lstick{\ket{\phi}_{\mathcal{Q}_1}} & \gate{\bar{O}_{\mathcal{Q}_1}} \gategroup{1}{2}{2}{2}{.7em}{--} & \multigate{4}{\text{QEC}^{*}} & \gate{\bar{O}_{\mathcal{Q}_1}} \gategroup{1}{4}{3}{4}{.7em}{--} & \multigate{4}{\text{QEC}^{*}} & \push{\dots} & & \gate{\bar{O}_{\mathcal{Q}_1}} \gategroup{1}{8}{5}{8}{.7em}{--} & \multigate{4}{\text{QEC}} & \qw \\
                \lstick{\mathcal{S}_0} & \ctrl{-1} & \ghost{\text{QEC}^{*}} & \qw & \ghost{\text{QEC}^{*}} & \dots & & \qw & \ghost{\text{QEC}} & \qw \\
                \lstick{\mathcal{S}_1} & \qw & \ghost{\text{QEC}^{*}} & \ctrl{-2} & \ghost{\text{QEC}^{*}} & \dots & & \qw & \ghost{\text{QEC}} & \qw \\
                \vdots & & \pureghost{\text{QEC}^{*}} & & \pureghost{\text{QEC}^{*}} & \ddots & & & \pureghost{\text{QEC}} & \\
                \lstick{\mathcal{S}_{a - 1}} & \qw & \ghost{\text{QEC}^{*}} & \qw & \ghost{\text{QEC}^{*}} & \dots & & \ctrl{-4} & \ghost{\text{QEC}} & \qw 
                \inputgroupv{3}{4}{6em}{1em}{\ket{\psi}_{\revBoth{GSCX}} \hspace{7em}}
            }
        }
    \end{subfigure}
    
    \caption{
    \revOne{Construction of the logical
    $\overline{CO}_{\revBoth{GSCX}_{a,b},\mathcal{Q}_1}$ gate.
    (a) Each qubit $s_j$ in a single subregister $\mathcal{S}_i$
    controls the corresponding physical gate $O_j$ comprising
    $\bar{O}_{\mathcal{Q}_1}
    = \prod_{j=0}^{b-1} O_j$,
    thereby performing
    $\overline{CO}_{\mathcal{S}_i,\mathcal{Q}_1}$
    transversally.
    Here, $d_j$ denotes the $j$th data qubit and $s_j$ denotes
    the $j$th qubit of $\mathcal{S}_i$.
    (b) The complete logical gate is obtained by applying the
    interaction in (a) sequentially for each subregister
    $\mathcal{S}_0,\ldots,\mathcal{S}_{a-1}$.
    Dashed boxes identify the interaction expanded in (a).
    Intermediate $\mathrm{QEC}^{*}$ rounds use the modified stabilizer
    of Eq.~\ref{eq:modified-stabilizers}, while the final
    $\mathrm{QEC}$ round uses the original stabilizers.
    }}
    \label{fig:circuits-gsch-controlled-flips}
\end{figure}

\revSelf{The fault-tolerant realization of our ancilla–data primitive is pieceable in the sense of Ref.~\cite{pieceableFt}.
The full logical interaction is decomposed into transversal subregister interactions separated by intermediate error correction. 
In contrast with previous pieceable constructions, however, the decomposition is determined by the structure of a fixed helper code rather than by the structure of the data code. 
This enables the same construction to target arbitrary stabilizer codes using only their logical Pauli representatives.}

After modifying the stabilizers, both the $\revBoth{GSCX}$ stabilizers and the data \revSelf{code} stabilizers \revOne{can be thought of as one large code.
However, each code can still be decoded independently. 
To see this, first note that the target code stabilizers remain unchanged and can thus act normally.
Any errors found on the target code can then be used to invert the measurement result of the modified stabilizer if they anticommute with it.
This reverts the stabilizer modification, which then allows the syndrome from the $\revBoth{GSCX}$ stabilizers to be decoded independently as well.
This is explained with more detail in Supplementary Note 2.4.
}

The number of correctable errors in each partition thus remains the same (we use ``partition'' to describe a quantum register tied to a code block).\ 
So, using a $\revBoth{GSCX}_{3,3}$ code targeting data encoded with a distance 3 code, the combined code can correct up to one arbitrary error in each partition.
This enables the correction of errors that have propagated from one partition to the other \revOne{(see Supplementary Note 2.5)}.\ 
An example along with further elaboration on the modified stabilizers can be found in Supplementary Note 2.1.

\subsection{Stabilizer Code-Generic Gates}\label{sec:scg}

Let the maximum number of physical gates between $\bar{X}_{\mathcal{Q}_1}$ and $\bar{Z}_{\mathcal{Q}_1}$ be $|\mathcal{L}|$ for some $\mathcal{Q}_1$ on which an ancilla register will target.
The following implementations use ancilla registers encoded with $\revBoth{GSCX}_{a,b}$, where $b \geq |\mathcal{L}|$.
$a$ and $b$ must be at least 3 in order to correct at least one arbitrary error, but in general one would likely want \revTwo{$a \geq d$, where $d$ is the distance of the $\mathcal{Q}_1$ code, and $b \geq |\bar{O}_{\mathcal{Q}_1}|$, where $\bar{O}_{\mathcal{Q}_1}$ is the weight of the representative of the logical operator $\bar{O}_{\mathcal{Q}_1}$}.

{\normalcolor The ancilla register will be encoded in $\ket{+}_{GSCX}$, which is equivalently done by preparing $\ket{0}_{GSC}$.
The circuits for Theorems~\ref{thm:h} and~\ref{thm:flip} are shown in Figs.~\ref{fig:hadamard-fault-tolerant} and~\ref{fig:controlled-fault-tolerant}, respectively, where we can fault-tolerantly perform the controlled quantum gates using Lemma~\ref{lem:cx}.
The proofs for the effect of the pre-measurement gates are shown in Supplementary Note 1.2
and Supplementary Note 1.3.
Note that data codes are not limited to single-logical-qubit QEC codes.
Consider, for example, a $\mathcal{Q}_2$ that refers to the second qubit of a code $\mathcal{C}$ encoding two logical qubits.
Then, $\bar{H}_{\mathcal{Q}_2}$ applied to $\ket{\psi}_{\mathcal{C}} = \alpha \ket{00}_{\mathcal{C}} + \beta \ket{01}_{\mathcal{C}} + \gamma \ket{10}_{\mathcal{C}} + \delta \ket{11}_{\mathcal{C}}$ results in the state $\alpha \ket{0+}_{\mathcal{C}} + \beta \ket{0-}_{\mathcal{C}} + \gamma \ket{1+}_{\mathcal{C}} + \delta \ket{1-}_{\mathcal{C}}$.

The Hadamard gate in Fig.~\ref{fig:controlled-fault-tolerant} can be done using Theorem~\ref{thm:h}.
Hence, we know how to perform all the quantum gates for these circuits fault-tolerantly.
The procedures are completed by measuring \revBoth{GSCX in the X-basis} and applying the appropriate corrections based on the measurement results.

\begin{theorem}\label{thm:h}
    \revOne{Given the initial state $\revBoth{\ket{+}_{GSCX}} \ket{\psi}_{\mathcal{Q}_1}$, the circuit shown in Fig.~\ref{fig:hadamard-fault-tolerant} fault-tolerantly performs an SCG Hadamard gate $\bar{H}_{\mathcal{Q}_1}$ such that 
    \begin{equation}
        \bar{H}_{\mathcal{Q}_1} \ket{\psi}_{\mathcal{Q}_1} = \frac{1}{\sqrt{2}} \left( (\alpha + \beta) \ket{0}_{\mathcal{Q}_1} + (\alpha - \beta) \ket{1}_{\mathcal{Q}_1} \right).
    \end{equation}}
\end{theorem}

\begin{theorem}\label{thm:flip}
    \revOne{Given the initial state $\revBoth{\ket{+}_{GSCX}} \ket{\psi}_{\mathcal{Q}_1} \ket{\phi}_{\mathcal{Q}_2}$, the circuit shown in Fig.~\ref{fig:controlled-fault-tolerant} fault-tolerantly performs an SCG $\overline{CO}_{\mathcal{Q}_1,\mathcal{Q}_2}$ gate, for any gate $O \in \{ X, Z \}$, such that 
    \begin{equation}
        \overline{CO}_{\mathcal{Q}_1,\mathcal{Q}_2} \ket{\psi}_{\mathcal{Q}_1} \ket{\phi}_{\mathcal{Q}_2} = \alpha \ket{0}_{\mathcal{Q}_1} \ket{\phi}_{\mathcal{Q}_2} + \beta \ket{1}_{\mathcal{Q}_1}(\bar{O}_{\mathcal{Q}_2} \ket{\phi}_{\mathcal{Q}_2}).
    \end{equation}}
\end{theorem}

We now have all the tools necessary to finish our SCG fault-tolerant universal gate set, which only needs an additional, fault-tolerant implementation of Z-rotations.
For this, we first encode an additional register in the logical zero state of some QEC code that has a fault-tolerant implementation of the desired rotation.
For example, we can use the Steane code for a $Z^{1/2} = S$ rotation or a [[15, 1, 3]] quantum Reed-Muller code for a $Z^{1/4} = T$ rotation~\cite{steaneCode,conversionSteaneReedMuller,triorthogonal,knill1996thresholdaccuracyquantumcomputation,reedMuller}, which have transversal implementations of the desired rotations.
Label this helper code, that has a fault-tolerant logical $\bar{Z}^{p}$ gate, as $RC$.

\revSelf{Two circuits performing Theorem~\ref{thm:z-fault-tolerant} via a $\bar{Z}^p$ measurement~\cite{magicStateNotAsCostly} are shown in Figs.~\ref{fig:t-fault-tolerant} and~\ref{fig:t-fault-tolerant-2}.
Fig.~\ref{fig:t-fault-tolerant} can be done entirely using gates described in this paper, but likely uses more qubits, depending on the implementation.
Alternatively, if a $\overline{Z_{\mathcal{Q}_1}Z_{RC}}$ measurement can be performed, Fig.~\ref{fig:t-fault-tolerant-2} can be done using the cheaper SCG Hadamard gate instead of an SCG controlled-$\bar{X}$ gate.}

\begin{theorem}\label{thm:z-fault-tolerant}
    \revOne{Given the initial state $\revBoth{\ket{+}_{GSCX}} \ket{\psi}_{\mathcal{Q}_1} \ket{0}_{RC}$, where $RC$ is a QEC code that has a fault-tolerant implementation of $\bar{Z}^{p} = \begin{bmatrix}
        1 & 0 \\ 0 & e^{i \pi p}
    \end{bmatrix}$, the circuits shown in Figs.~\ref{fig:t-fault-tolerant} and \ref{fig:t-fault-tolerant-2} each fault-tolerantly perform an SCG gate $\left( \bar{Z}^{p} \right)_{\mathcal{Q}_1}$, such that 
    \begin{equation}
        \left( \bar{Z}^{p} \right)_{\mathcal{Q}_1} \ket{\psi}_{\mathcal{Q}_1} = \alpha \ket{0}_{\mathcal{Q}_1} + e^{i \pi p}\beta \ket{1}_{\mathcal{Q}_1}.
    \end{equation}}
\end{theorem}

\begin{figure}
    \centering
    \captionsetup{justification=raggedright,singlelinecheck=false}
    \begin{subfigure}{\textwidth}
        \centering
        \caption{}
        \label{fig:hadamard-fault-tolerant}
        \centerline{
            \Qcircuit @C=1em @R=.8em {
                \lstick{\ket{\psi}_{\mathcal{Q}_1}} & \gate{\bar{X}} & \gate{\bar{Z}} & \qw & \gate{\bar{X}} & \gate{\bar{Z}} & \qw &  \rstick{\bar{H}_{\mathcal{Q}_1} \ket{\psi}_{\mathcal{Q}_1}} \\
                \lstick{\ket{+}_{GSCX}} & \ctrl{-1} & \ctrl{-1} & \measureD{\bar{X}} & \cctrl{-1} & \cctrlo{-1} &  &  \\
                \protected\gategroup{1}{2}{2}{2}{.8em}{--}
                \protected\gategroup{1}{3}{2}{3}{.8em}{--}
            }
        }
    \end{subfigure}
    \hfill
    \begin{subfigure}{\textwidth}
        \centering
        \caption{}
        \label{fig:controlled-fault-tolerant}
        \centerline{
            \Qcircuit @C=1em @R=.8em {
                \lstick{\ket{\psi}_{\mathcal{Q}_1}} & \gate{\bar{Z}} & \qw & \qw & \qw & \gate{\bar{Z}} & \qw & \hspace{9em} \raisebox{-3em}{$\overline{CO}_{\mathcal{Q}_1, \mathcal{Q}_2} \ket{\psi}_{\mathcal{Q}_1} \ket{\phi}_{\mathcal{Q}_2}$} \\
                \lstick{\ket{\phi}_{\mathcal{Q}_2}} & \qw & \qw & \gate{\bar{O}} & \qw & \qw & \qw \gategroup{1}{7}{2}{7}{.8em}{\}} & \\
                \lstick{\ket{+}_{GSCX}} & \ctrl{-2} & \gate{\bar{H}} & \ctrl{-1} & \measureD{\bar{X}} & \cctrl{-2} &  &  \\
                \protected\gategroup{1}{2}{3}{2}{.8em}{--}
                \protected\gategroup{3}{3}{3}{3}{.8em}{.}
                \protected\gategroup{2}{4}{3}{4}{.8em}{--}
            }
        }
    \end{subfigure}
    \hfill
    \begin{subfigure}{\textwidth}
        \centering
        \caption{}
        \centerline{
            \Qcircuit @C=1em @R=.8em {
                \lstick{\ket{\psi}_{Q1}} & \ctrl{1} & \qw & \qw & \gate{\bar{Z}} & \qw & \rstick{\left(\bar{Z}^p\right)_{\mathcal{Q}_1} \ket{\psi}_{\mathcal{Q}_1}} \\
                \lstick{\ket{0}_{RC}} & \targ & \gate{\bar{Z}^p} & \measureD{\bar{X}} & \cctrl{-1} & & \\
                \protected\gategroup{1}{2}{2}{2}{.8em}{.}
            }
        }
        \label{fig:t-fault-tolerant}
    \end{subfigure}
    \hfill
    \begin{subfigure}{\textwidth}
        \centering
        \caption{}
        \label{fig:t-fault-tolerant-2}
        \centerline{
            \Qcircuit @C=1em @R=.8em {
                \lstick{\ket{\psi}_{\mathcal{Q}_1}} & \qw & \multimeasureD{2}{\overline{ZZ}} & \qw & \qw & \qw & \gate{\bar{Z}} & \qw & \rstick{\left(\bar{Z}^p\right)_{\mathcal{Q}_1} \ket{\psi}_{\mathcal{Q}_1}} \\
                 &  & \pureghost{\overline{ZZ}} & \cctrl{1} &  &  &  &  &  \\
                \lstick{\ket{0}_{RC}} & \gate{\bar{H}} & \ghost{\overline{ZZ}} & \gate{\bar{X}} & \gate{\bar{Z}^p} & \measureD{\bar{X}} & \cctrl{-2} &  & \\
                \protected\gategroup{3}{2}{3}{2}{.8em}{.}
            }
        }
    \end{subfigure}
    \caption{Circuit representation of fault-tolerant SCG (a) $\bar{H}_{\mathcal{Q}_1}$, (b) $\overline{CO}_{\mathcal{Q}_1,\mathcal{Q}_2}$, and (c)\revSelf{/(d)} $\left(\bar{Z}^p\right)_{\mathcal{Q}_1}$ gates.
    Dashed boxes represent $\bar{X}/\bar{Z}$ gates controlled by a $\revBoth{GSCX}$ register, whose process is described with Lemma~\ref{lem:cx} and shown in Fig.~\ref{fig:gsch-sequence}.
    Dotted boxes represent SCG gates.
    The $\overline{ZZ}$ measurement in (\revSelf{d}) measures the parity of the $\bar{Z}$ observables in each register.
    The $RC$ subscript denotes a code that has a fault-tolerant implementation of a desired logical $\bar{Z}^p$ gate.
    \revSelf{All single-register measurements are done in the X-basis.}}
\end{figure}
}

\subsection{Validation}

\subsubsection{Logical Error Rate}

\begin{figure*}
    \centering
    \includegraphics[width=.81\textwidth]{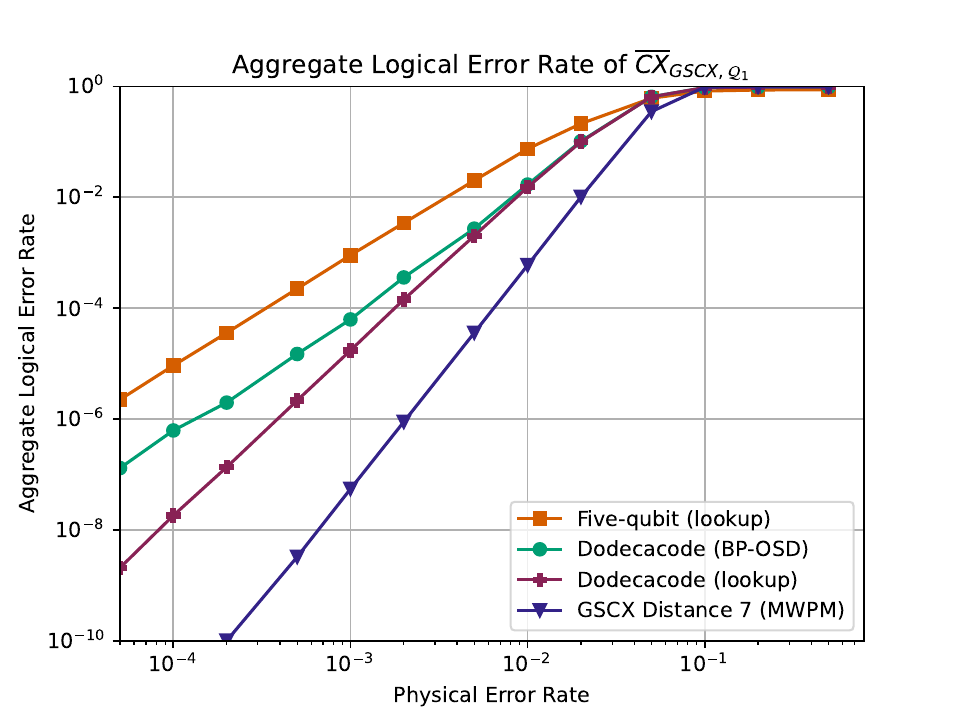}
    \caption{\revOne{The aggregate LER under phenomenological noise of $\overline{CX}_{\revBoth{GSCX},\mathcal{Q}_1}$ acting on $\ket{+}_{GSCX}\ket{0}_{Q1}$.
    The parentheses in the labels name the decoder used for the target code.}
    }
    \label{fig:qec-graphs}
\end{figure*}

The \revSelf{results of the} simulations described in Sec.~\ref{sec:methodology} are shown in Fig.~\ref{fig:qec-graphs}.
\revOne{
This figure shows the aggregate logical error rate (LER) of the logical $\overline{CX}_{\revBoth{GSCX},\mathcal{Q}_1}$ \revSelf{gate} under phenomenological noise across all subregisters in a Z-basis experiment.
\revSelf{For the exact lookup decoders and MWPM decoder evaluated here, the LER exhibits the expected distance-dependent suppression. 
In particular, for constituent codes of distance $d$, these results are consistent with $O(p^{t+1})$, where $t=\lfloor(d-1)/2\rfloor$.
For the dodecacode, BP-OSD closely matches the exact lookup decoder at higher physical error rates but exhibits weaker suppression as the physical error rate decreases.
Since the exact lookup decoder for the same code retains the expected distance-dependent scaling, this difference reflects decoder-dependent performance rather than a reduction in the error-correcting capability of the modified-stabilizer construction.
Thus, the simulations support that the modified-stabilizer construction itself preserves the expected $O(p^{t+1})$ error-suppression order of the constituent codes.
Since the logical $\overline{CX}_{GSCX,\mathcal{Q}_1}$ and $\overline{CZ}_{GSCX,\mathcal{Q}_1}$ gates have the same fault-tolerant construction, and each SCG gate contains only a constant number of such primitives, the $O(p^{t+1})$ LER scaling is preserved for the complete SCG procedures.}
}

\subsubsection{SCG Gate Effects}\label{sec:validation-effects}

\revSelf{With the preservation of LER demonstrated, we noiselessly simulated the full gate sequences of all the SCG gates described in Sec.~\ref{sec:scg} to verify that they output the expected state vectors.}
These simulations acted on randomly generated states encoded with various QEC codes, and the resulting state vectors precisely matched the expected theoretical outcomes.
These noiseless simulations support that the proposed protocols implement the correct logical transformations, thereby \revSelf{demonstrating the effects of} the SCG gate set $\{\bar{H}, \overline{CX}, \bar{T}\}$.

For example, acting on the four-qubit code ($[[4, 2, 2]]$)~\cite{four-qubit}, where the logical input state is $\ket{\psi}_{\mathcal{Q}_1} = \alpha \ket{00}_{\mathcal{Q}_1} + \beta \ket{01}_{\mathcal{Q}_1} + \gamma \ket{10}_{\mathcal{Q}_1} + \delta \ket{11}_{\mathcal{Q}_1}$ with randomly generated $\alpha, \beta, \gamma,$ and $\delta$, the SCG gates yielded the states given in Table~\ref{tab:stabilizer-generic-gate-effects}.
\begin{table}
    \centering
    \caption{\revSelf{SCG gate effects on the $\mathcal{C} =[[4, 2, 2]]$ code acting on initial state $\alpha \ket{00}_\mathcal{C} + \beta \ket{01}_\mathcal{C} + \gamma \ket{10}_\mathcal{C} + \delta \ket{11}_\mathcal{C}$.}}
    \label{tab:stabilizer-generic-gate-effects}
    \revSelf{\begin{tabular}{Sc Sc}
        \toprule
        \textbf{Gate} & \textbf{Output} \\
        \midrule
        $\bar{H}_{\mathcal{Q}_2}$ & $\frac{1}{\sqrt{2}}(\alpha + \beta) \ket{00}_{\mathcal{C}}$ \\
         & $+ \frac{1}{\sqrt{2}}(\alpha - \beta) \ket{01}_{\mathcal{C}}$ \\
         & $ + \frac{1}{\sqrt{2}}(\gamma + \delta) \ket{10}_{\mathcal{C}}$ \\
          & $+ \frac{1}{\sqrt{2}}(\gamma - \delta) \ket{11}_{\mathcal{C}}$ \\
        \hline
        $\overline{CX}_{\mathcal{Q}_1,\mathcal{Q}_2}$ & $\alpha \ket{00}_{\mathcal{C}} + \beta \ket{01}_{\mathcal{C}}$ \\
         & $+ \delta \ket{10}_{\mathcal{C}} + \gamma \ket{11}_{\mathcal{C}}$ \\
        \hline
        $\bar{T}_{\mathcal{Q}_2}$ & $\alpha \ket{00}_{\mathcal{C}} + e^{i \pi / 4} \beta \ket{01}_{\mathcal{C}}$ \\
         & $+ \gamma \ket{10}_{\mathcal{C}} + e^{i \pi / 4} \delta \ket{11}_{\mathcal{C}}$ \\
        \bottomrule
    \end{tabular}}
\end{table}
Further experiments included the implementation of the Deutsch-Jozsa algorithm~\cite{Deutsch-Jozsa} using $GSC_{3,3}$ for two logical data qubits and one logical qubit for phase kickback. 
As expected, only the all-zero state was measured when the oracle was constant, and anything but the all-zero state was measured when the oracle was balanced.

\section{Discussion}

In this work, we have demonstrated an SCG framework for universal fault-tolerant quantum computation. 
Our novel, ancilla-mediated approach effectively circumvents the Eastin-Knill theorem~\cite{eastinKnill} without relying on constricting code-specific properties abundant in previous methods. 
Our results confirm that universal fault tolerance can be achieved for \emph{any} stabilizer code, provided that the appropriate $\revBoth{GSCX}$ and an appropriate code with a fault-tolerant logical $\bar{T}$ gate can be prepared.
Furthermore, the data remains encoded in the same physical space, so no modification to the data codes or registers is required.
\revOne{By decoupling the logical gate implementation from the underlying code structure, our method also enables heterogeneous logical gates, rendering any stabilizer code capable of interacting with any other stabilizer code.}

\revOne{One future application is, with the general convergence on distributed quantum computing~\cite{distributedQc}, it may be beneficial to distribute computations across different qubit modalities.}
Since certain hardware may be better suited for certain QEC codes, such as superconducting hardware for surface codes~\cite{surfaceCodeSuperconducting} or neutral atom hardware for the quantum low-density parity-check (QLDPC) code family~\cite{neutralAtomQLDPC}, heterogeneous communication between different stabilizer codes would facilitate distribution between different qubit modalities.
\revTwo{Notably, however, the required $GSCX$ subregister size $b$ is at least the weight of the chosen representative of the target logical $\overline{X}$ or $\overline{Z}$ operator.
For QLDPC codes, finding sufficiently low-weight representatives may therefore affect the practicality of applying the construction.}

Notably, these SCG gate protocols are not necessarily all required to construct a universal gate set for a particular code.
These protocols serve as generic methods if specialized methods are unavailable.
For example, lattice or QLDPC surgery already provides low-cost logical controlled-NOT gates for viable codes~\cite{latticeSurgery,qldpcSurgery}.
Hence, a computation using, say, the surface code may opt to use lattice surgery for logical controlled-NOT gates but still may find our SCG $T$ gate useful compared to magic state distillation.
In other words, since each of our stabilizer code-generic gates do not rely on or change the underlying structure of any code, the computation is free to use them as-needed to optimize any part of the broader system.
\revOne{They provide generic alternatives where specialized techniques are unavailable.}

\revOne{A distinguishing feature of the framework is its ability to mediate logical interactions between arbitrary stabilizer data codes, whereas magic state distillation is commonly used to extend the gate set available to a single code.}
This strategy's standard implementations rely on distillation factories, which impose significant spacetime overheads and are by nature \revSelf{probabilistic}.
That is, there is some probability of failure, and the process must be repeated indefinitely until a success occurs.
In contrast, our protocol operates deterministically with a linear overhead of QEC rounds by code distance.
A deeper analysis of tradeoffs between these two methods, investigating tradeoffs between runtime, number of qubits, and number of gates, is also left for future work.

As detailed in Supplementary Note 3,
the qubit overhead for our protocol is primarily due to the ancillary registers required for the $\revBoth{GSCX}$ states.
While the physical (qubits, gates) and temporal scaling is polynomial, this scheme can require high-weight stabilizers, especially the modified stabilizers.
Thus, measuring these high-weight stabilizers, for example by preparing large cat states, can drastically increase the time complexity and noise susceptibility.
Other methods of measurement, such as homomorphic measurement~\cite{homomorphicMeasurements,homomorphicComputation}, may be incorporated or developed in the future to reduce this cost.

The scheme presented here, utilizing $\revBoth{GSCX}$ for the ancilla, will likely not be \revOne{directly} useful \revOne{under circuit-level noise} due to the high-weight stabilizers leading to a \revSelf{low} or nonexistent threshold.
However, this work explains the fundamental principles of a complete, fault-tolerant, universal gate set using modified stabilizers.
\revOne{Since a circuit-level LER and threshold would depend highly on the choice of codes, measurement schemes, decoders, and other implementation details,
specific implementations using various codes and hardware may be developed.}
For example, it may be possible that the surface code can be used \revSelf{instead of $\revBoth{GSCX}$} for the ancilla register, and a 3D color code~\cite{gaugeColorCodes,costDistillationVsColorCodeSwitching} can be used for the $T$ rotation.
As both of these have been shown to have \revSelf{an achievable threshold}, the principles in this paper being applied with these codes may prove useful.

Other schemes may be developed to reduce this weight, for example by using gauges, efficient codes, or single-shot, multi-target gates~\cite{multiTargetryGates}, but we leave this to future work.
\revOne{Ultimately, this framework could support communication within hybrid quantum architectures in which, for example, high-threshold codes serve as memory and high-rate codes serve as processing units.
This could support modular quantum architectures in which different stabilizer codes serve different roles while reducing the need to develop a separate universal gate construction for each code.}

\section{Methods}\label{sec:methodology}

To substantiate and provide numerical evidence of the proposed fault-tolerant protocols, we conducted numerical simulation experiments divided into two main parts: (1) verification of the error correction capabilities during the execution of the full logical $\overline{CX}$ gate controlled by \revBoth{GSCX} (see Sec.~\ref{sec:cx-by-cpc}) and (2) confirmation of the unitary effects of the SCG gates. 
To achieve this second objective, we utilized the \textit{cirq} Python package to simulate the gate sequences and verify the resulting state vectors, as mentioned in Sec.
~\ref{sec:validation-effects}.

\revTwo{The Deutsch-Jozsa algorithm was simulated to support the validity of the SCG Hadamard and $\overline{CX}$ gates.
Three $GSC_{3,3}$ qubits were initialized in their logical $\ket{0}$ states, two for the data and one for the ancilla used for phase kickback.
Two oracles were created, one balanced and one constant.
The constant oracle acted as an identity operator, and the balanced performed two SCG $\overline{CX}$ gates, one for each data qubit targeting the ancilla.
After flipping the logical ancilla qubit using its logical $\bar{X}$ gate, SCG Hadamard gates were performed on all logical qubits.
The given oracle was performed, SCG Hadamards were again performed on all logical qubits, and the ancilla was flipped back to its initial state.
Finally, the data qubits were measured.
Each oracle was tested with ten random seeds, which we believe to be sufficient to reliably account for the possibilities of only two data qubits.
With a 100\% success rate, the measurements resulted in all zeros in every case when the constant oracle was used, and at least one $1$ was measured in every case when the balanced oracle was used.}

Error correction was validated using the \textit{stim}~\cite{stim} \revSelf{and \textit{sinter} Python packages}.
The only additional QEC proposed in this paper is applied after each transversal logical controlled-$\bar{X}/\bar{Z}$ gate controlled by a $\revBoth{GSCX}$ subregister.
Hence, these simulations assume fault-tolerant encoding and measurement procedures to isolate the capabilities of the modified stabilizers used in this process.

{\normalcolor
\paragraph{Transversal Step Simulation} To test error correction after a specific transversal operation, $\overline{CX}_{\mathcal{S}_i, \mathcal{Q}_1}$ (for $i \in \{0, \dots, d-1\}$), all preceding transversal gates $\overline{CX}_{\mathcal{S}_j, \mathcal{Q}_1}$ with $j < i$ were performed noiselessly. 
Single-qubit depolarizing noise with probability $p$ was applied to all data qubits immediately prior to the $\overline{CX}_{\mathcal{S}_i, \mathcal{Q}_1}$ gate.
Immediately afterwards, two-qubit depolarizing noise with probability $p$ was applied to each interacting qubit pair in $\overline{CX}_{\mathcal{S}_i, \mathcal{Q}_1}$, and single-qubit depolarizing noise was applied to idle data qubits.
Error correction was then executed with the modified stabilizer given by Eq.~\eqref{eq:modified-stabilizers}.

\paragraph{Error Correction Rounds} Error correction was simulated using a phenomenological noise model. 
Between $d$ noisy syndrome-extraction rounds, single-qubit depolarizing noise with probability $p$ was applied to all data qubits, and a measurement readout error probability of $p$ was applied to the syndrome extraction measurements. 
Adopting a common modeling choice for quantum-memory readout in phenomenological-noise studies, we append an ideal final syndrome-extraction round and perfect final data-qubit measurements to serve as a clean readout boundary~\cite{kuo2023generalized,huang2026high,costDistillationVsColorCodeSwitching}.

\paragraph{Observable} For the $i$th subregister $\mathcal{S}_i$, the logical observable was defined over the combined system as the target logical $\bar{Z}_{\mathcal{Q}_1}$ combined with the $Z$ observable on the first qubit of each active control subregister $\mathcal{S}_j$ for $j \leq i$.
Since these subregisters are entangled with $\mathcal{Q}_1$, their combined parity under noiseless conditions yields an eigenvalue of $+1$.

\paragraph{Two-Stage Partition Decoding} Because the target code remains unmodified, and the modified stabilizer can be returned to its original state in post-processing,
decoding was performed via a two-stage partition decoder (see Supplementary Note 2.4):
\begin{enumerate}
    \item \textit{Target Partition:} The target code's syndrome was decoded without any adjustment from the $\overline{CX}_{\revBoth{GSCX},\mathcal{Q}_1}$ procedure.
    An exact single-error lookup table was used for the five-qubit code ($[[5,1,3]]$).
    Belief Propagation with Ordered Statistics Decoding (BP-OSD)~\cite{Roffe2020,Panteleev2021} (min-sum BP with scaling factor $0.625$, capped at 100 iterations, and order-40 combination sweep) as well as an exact weight $\leq 2$ lookup table was used for the dodecacode ($[[11,1,5]]$).
    Minimum-Weight Perfect Matching (MWPM) via \textit{PyMatching}~\cite{higgott2021pymatchingpythonpackagedecoding} was used for the \revBoth{GSCX} target code.
    \item \textit{Stabilizer Unmodification:} The target decoder extracted per-round error predictions. 
    For any round with a predicted target error that anticommuted with the modified stabilizer, the modified stabilizer's measurement result was inverted.
    We call this ``unmodifying'' the modified stabilizer.
    \item \textit{Control Partition:} The ``unmodified'' stabilizer and its result decouple the control code syndrome from the target partition. 
    The adjusted syndrome was then decoded using MWPM on the \revBoth{GSCX} detector graph.
    \item \textit{Combined Prediction:} The final observable prediction was computed as the modulo-2 sum of the target and control observable predictions.
\end{enumerate}

\paragraph{Aggregate Logical Error Rate}
To evaluate the overall fault tolerance of the complete logical $\overline{CX}_{\revBoth{GSCX},\mathcal{Q}_1}$ gate, the total logical error rate ($\text{LER}_{\text{total}}$) was aggregated over all $d$ transversal subregister stages:
\begin{equation}
    \text{LER}_{\text{total}}(p) = 1 - \prod_{i=0}^{d-1} \bigl(1 - \text{LER}_i(p)\bigr),
\end{equation}
where $\text{LER}_i(p)$ is the logical error rate of the $i$-th transversal step.
}

\section*{Data Availability}
All data generated or analyzed during this study are included in this published article (and its Supplementary Information files).
The datasets used and/or analysed during the current study available from the corresponding author on reasonable request.

\section*{Code Availability} 
The code used to simulate the protocols and generate the data in this study is available in the CU-Quantum/stabilizer-code-generic-ftqc repository at https://doi.org/10.5281/zenodo.22149962~\cite{zenodo}.

\section*{Acknowledgements}
This work utilized the Alpine high performance computing resource at the University of Colorado Boulder. Alpine is jointly funded by the University of Colorado Boulder, the University of Colorado Anschutz, Colorado State University, and the National Science Foundation (award 2201538).

This research was also supported (in part) by the National Science Foundation under award 2544544.

Diagrams were created in Lucid (lucid.co).

\section*{Author contributions statement}

N.J.P. conceived the project, developed the protocols, and performed the experiments. 
R.A. supervised the work. 
Both authors wrote and reviewed the manuscript.

\section*{Competing Interests}
The authors declare no competing interests.

\clearpage
\renewcommand{\refname}{Supplementary References}


\setcounter{table}{0}
\renewcommand{\tablename}{Supplementary Table}
\renewcommand{\thetable}{\arabic{table}} 

\setcounter{section}{0}
\setcounter{figure}{0}
\setcounter{table}{0}
\setcounter{equation}{0}
\setcounter{page}{1} 

\renewcommand{\thesection}{Supplementary Note \arabic{section}}
\renewcommand{\thefigure}{S\arabic{figure}}
\renewcommand{\thetable}{S\arabic{table}}
\renewcommand{\theequation}{S\arabic{equation}}

\begin{center}
    \vspace*{1cm}
    {\huge \textbf{Supplementary Information for:}} \\[0.5cm]
    {\LARGE Stabilizer Code-Generic Universal Fault-Tolerant Quantum Computation} \\[0.5cm]
    {\large Nicholas J.C. Papadopoulos and Ramin Ayanzadeh} \\
    \vspace*{1cm}
\end{center}


\section{Equations and Examples}

The following sections use these shorthand notations:
\begin{equation}
    \begin{aligned}
        \tilde{\ket{\psi}}_{\mathcal{Q}_1} &= \bar{X}_{\mathcal{Q}_1}\ket{\psi}_{\mathcal{Q}_1} \\
        &= \beta \ket{0}_{\mathcal{Q}_1} + \alpha \ket{1}_{\mathcal{Q}_1} \\
        \hat{\ket{\psi}}_{\mathcal{Q}_1} &= \bar{Z}_{\mathcal{Q}_1}\ket{\psi}_{\mathcal{Q}_1} \\
        &= \alpha \ket{0}_{\mathcal{Q}_1} - \beta \ket{1}_{\mathcal{Q}_1} \\
        \hat{\tilde{\ket{\psi}}}_{\mathcal{Q}_1} &= \bar{Z}_{\mathcal{Q}_1}\bar{X}_{\mathcal{Q}_1}\ket{\psi}_{\mathcal{Q}_1} \\
        &= \beta \ket{0}_{\mathcal{Q}_1} - \alpha \ket{1}_{\mathcal{Q}_1} \\
    \end{aligned}
\end{equation}

\subsection{Helper Code Examples}\label{sec:helper-code-examples}

Table~\ref{tab:mcc-example-3-4} shows the stabilizers for $GSC_{3,4}$, while Eq.~\eqref{eq:cpc-example-3-4} shows the logical states of \revBoth{its X-basis}.
\begin{table}
  \centering
  \caption{The stabilizers for $GSC$ with $3$ cat states and $4$ qubits per cat state ($GSC_{3,4}$), having generators $g_i$ and logical operators $\bar{Z}$ and $\bar{X}$.}
  \label{tab:mcc-example-3-4}
  \begin{tabular}{cc}
    \toprule
    \textbf{Name} & \textbf{Operator} \\
    \midrule
    $g_0$ & $Z_0 Z_1$ \\
    $g_1$ & $Z_1 Z_2$ \\
    $g_2$ & $Z_2 Z_3$ \\
    $g_3$ & $Z_4 Z_5$ \\
    $g_4$ & $Z_5 Z_6$ \\
    $g_5$ & $Z_6 Z_7$ \\
    $g_6$ & $Z_8 Z_9$ \\ 
    $g_7$ & $Z_9 Z_{10}$ \\
    $g_8$ & $Z_{10} Z_{11}$ \\
    $g_9$ & $X_0 X_1 \dots X_7$ \\
    $g_{\revSelf{10}}$ & $X_4 X_5 \dots X_{11}$ \\
    $\bar{Z}$ & $X_0 X_1 X_2 X_3$ \\
    $\bar{X}$ & $Z_0 Z_4 Z_8$ \\
  \bottomrule
\end{tabular}
\end{table}

\begin{equation}\label{eq:cpc-example-3-4}
    \begin{aligned}
        \ket{0}_{\revBoth{GSCX}_{3,4}} &= \frac{1}{2} \left( \ket{0..0..0..} + \ket{0..1..1..} + \ket{1..0..1..} + \ket{1..1..0..} \right) \\
        \ket{1}_{\revBoth{GSCX}_{3,4}} &= \frac{1}{2} \left( \ket{0..0..1..} + \ket{0..1..0..} + \ket{1..0..0..} + \ket{1..1..1..} \right)
    \end{aligned}
\end{equation}
where $\ket{a..b..c..}$ is shorthand for $\ket{a}^{\otimes 4} \ket{b}^{\otimes 4} \ket{c}^{\otimes 4}$.

\subsection{SCG Hadamard}\label{sec:hadamard-single-ancilla-expanded}
The following details the action of the \revSelf{pre-measurement} operations of the SCG Hadamard gate.

\begin{equation}\label{eq:hadamard-single-ancilla-expanded}
 \revBoth{\begin{aligned}
    &\quad \overline{CZ}_{GSCX, \mathcal{Q}_1} \overline{CX}_{GSCX, \mathcal{Q}_1} \left( \ket{+}_{GSCX} \ket{\psi}_{\mathcal{Q}_1} \right) \\
    &= \frac{1}{\sqrt{2}} \overline{CZ}_{GSCX, \mathcal{Q}_1} \left( \ket{0}_{GSCX} \ket{\psi}_{\mathcal{Q}_1} + \ket{1}_{GSCX} \tilde{\ket{\psi}}_{\mathcal{Q}_1} \right) \\
    &= \frac{1}{\sqrt{2}} \left( \ket{0}_{GSCX} \ket{\psi}_{\mathcal{Q}_1} + \ket{1}_{GSCX} \hat{\tilde{\ket{\psi}}}_{\mathcal{Q}_1} \right) \\
    &= \frac{1}{2} \left( \left( \ket{+}_{GSCX} + \ket{-}_{GSCX} \right) \ket{\psi}_{\mathcal{Q}_1} + \left( \ket{+}_{GSCX} - \ket{-}_{GSCX} \right) \hat{\tilde{\ket{\psi}}}_{\mathcal{Q}_1} \right) \\
    &= \frac{1}{2} \left( \ket{+}_{GSCX} \left( \ket{\psi}_{\mathcal{Q}_1} + \hat{\tilde{\ket{\psi}}}_{\mathcal{Q}_1} \right) + \ket{-}_{GSCX} \left( \ket{\psi}_{\mathcal{Q}_1} - \hat{\tilde{\ket{\psi}}}_{\mathcal{Q}_1} \right) \right) \\
    &= \text{  } \frac{1}{2} \left( \ket{+}_{GSCX} \left( \alpha \ket{0}_{\mathcal{Q}_1} + \beta \ket{1}_{\mathcal{Q}_1} + \beta \ket{0}_{\mathcal{Q}_1} - \alpha \ket{1}_{\mathcal{Q}_1} \right) \right) \\
    &\quad + \frac{1}{2}  \left( \ket{-}_{GSCX} \left( \alpha \ket{0}_{\mathcal{Q}_1} + \beta \ket{1}_{\mathcal{Q}_1} - \beta \ket{0}_{\mathcal{Q}_1} + \alpha \ket{1}_{\mathcal{Q}_1} \right) \right) \\
    &= \text{  } \frac{1}{\sqrt{2}} \ket{+}_{GSCX} \left( \frac{1}{\sqrt{2}} \left( \left( \alpha + \beta \right) \ket{0}_{\mathcal{Q}_1} + \left( \beta - \alpha \right) \ket{1}_{\mathcal{Q}_1} \right) \right) \\
    &\quad + \frac{1}{\sqrt{2}} \ket{-}_{GSCX} \left( \frac{1}{\sqrt{2}} \left( \left( \alpha - \beta \right) \ket{0}_{\mathcal{Q}_1} + \left( \alpha + \beta \right) \ket{1}_{\mathcal{Q}_1} \right) \right) \\
 \end{aligned}}
\end{equation}

\subsection{SCG Controlled-$\bar{X}/\bar{Z}$}\label{sec:controlled-op-single-qubit-expanded}
The following details the action of the \revSelf{pre-measurement} operations of the SCG \revSelf{controlled-}$\bar{X}/\bar{Z}$ gate.

\begin{equation}\label{eq:controlled-op-single-qubit-expanded}
    \revBoth{\begin{aligned}
        &\quad \overline{CO}_{GSCX,\mathcal{Q}_2} H_{GSCX} \overline{CZ}_{GSCX,\mathcal{Q}_1} \left( \ket{+}_{GSCX} \ket{\psi}_{\mathcal{Q}_1} \ket{\phi}_{\mathcal{Q}_2} \right) \\
        &= \frac{1}{\sqrt{2}} \overline{CO}_{GSCX,\mathcal{Q}_2} H_{GSCX} \left(\ket{0}_{GSCX}\ket{\psi}_{\mathcal{Q}_1} + \ket{1}_{GSCX}\hat{\ket{\psi}}_{\mathcal{Q}_1}\right) \ket{\phi}_{\mathcal{Q}_2} \\
        &= \frac{1}{\sqrt{2}} \overline{CO}_{GSCX,\mathcal{Q}_2} \left(\ket{+}_{GSCX}\ket{\psi}_{\mathcal{Q}_1} + \ket{-}_{GSCX}\hat{\ket{\psi}}_{\mathcal{Q}_1}\right) \ket{\phi}_{\mathcal{Q}_2} \\
        &= \frac{1}{2} \overline{CO}_{GSCX,\mathcal{Q}_2} \left( \ket{0}_{GSCX} \left(\ket{\psi}_{\mathcal{Q}_1} + \hat{\ket{\psi}}_{\mathcal{Q}_1}\right) + \ket{1}_{GSCX} \left(\ket{\psi}_{\mathcal{Q}_1} - \hat{\ket{\psi}}_{\mathcal{Q}_1}\right) \right) \ket{\phi}_{\mathcal{Q}_2} \\
        &= \overline{CO}_{GSCX,\mathcal{Q}_2} (\alpha \ket{0}_{GSCX} \ket{0}_{\mathcal{Q}_1} + \beta \ket{1}_{GSCX} \ket{1}_{\mathcal{Q}_1}) \ket{\phi}_{\mathcal{Q}_2} \\
        &= \left( \alpha \ket{0}_{GSCX} \ket{0}_{\mathcal{Q}_1} \ket{\phi}_{\mathcal{Q}_2} + \beta \ket{1}_{GSCX} \ket{1}_{\mathcal{Q}_1} \left(\bar{O}_{\mathcal{Q}_2} \ket{\phi}_{\mathcal{Q}_2}\right) \right) \\
        &= \text{ } \frac{1}{\sqrt{2}} \left( \alpha (\ket{+}_{GSCX} + \ket{-}_{GSCX}) \ket{0}_{\mathcal{Q}_1} \ket{\phi}_{\mathcal{Q}_2} \right) \\
        &\quad + \frac{1}{\sqrt{2}} \left( \beta (\ket{+}_{GSCX} - \ket{-}_{GSCX}) \ket{1}_{\mathcal{Q}_1} \left(\bar{O}_{\mathcal{Q}_2} \ket{\phi}_{\mathcal{Q}_2}\right) \right) \\
        &= \text{ } \frac{1}{\sqrt{2}} \ket{+}_{GSCX} \left( \alpha \ket{0}_{\mathcal{Q}_1} \ket{\phi}_{\mathcal{Q}_2} + \beta \ket{1}_{\mathcal{Q}_1} \left(\bar{O}_{\mathcal{Q}_2} \ket{\phi}_{\mathcal{Q}_2}\right) \right) \\
        &\quad + \frac{1}{\sqrt{2}} \ket{-}_{GSCX} \left( \alpha \ket{0}_{\mathcal{Q}_1} \ket{\phi}_{\mathcal{Q}_2} - \beta \ket{1}_{\mathcal{Q}_1} \left(\bar{O}_{\mathcal{Q}_2} \ket{\phi}_{\mathcal{Q}_2}\right) \right)
    \end{aligned}}
\end{equation}

\section{$\bar{X}/\bar{Z}$ Controlled by $\revBoth{GSCX}$, Modified Stabilizers}\label{sec:modified-stabilizers}

\revOne{\subsection{Stabilizer Modification to Keep the State Within the Codespace}}
\revSelf{The modified stabilizers after a $\overline{CO}_{\mathcal{S}_i,\mathcal{Q}_1}$ gate appear inherently after conjugating the $GSCX$ stabilizers with $\overline{CO}_{\mathcal{S}_j,\mathcal{Q}_1}$ gates, for all $j \leq i$.
The target code stabilizers do not need to be modified because only a logical operation is ever applied to them, leaving the target codespace intact.
Below, we provide further intuition for why the modified stabilizers keep the state within the codespace.}

Given $\ket{\phi}_{\revBoth{GSCX}_{a,b}} \ket{\psi}_{\mathcal{Q}_1}$, each subspace $\revSelf{\alpha}\ket{0}_{\revBoth{GSCX}_{a,b}}\ket{\psi}_{\mathcal{Q}_1}$ and $\revSelf{\beta}\ket{1}_{\revBoth{GSCX}_{a,b}}\ket{\psi}_{\mathcal{Q}_1}$ will be further separated into two subspaces after each $\overline{CO}_{\mathcal{S}_i,\mathcal{Q}_1}$ for sequential $\mathcal{S}_i$ in $\revBoth{GSCX}_{a,b}$.
This section aims to show that
\begin{equation}
    \begin{aligned}
        g_x'(i, j) = g_x(j) \left( \bar{O}_{\mathcal{Q}_1} \right)^{\delta_{ij}}
    \end{aligned}
\end{equation}
serves as the modified $j$th X-stabilizer, after performing $\overline{CO}_{\mathcal{S}_i,\mathcal{Q}_1}$, \revSelf{that maintains the codespace.}

First recognize that, for any computational basis state $\ket{x}_{\revBoth{GSCX}_{a,b}}$, $\overline{CO}_{\mathcal{S}_i,\mathcal{Q}_1}$ splits the basis states of $\ket{x}_{\revBoth{GSCX}_{a,b}}$ by whether an even number of the first $i\revSelf{+1}$ cat states are in $\revSelf{\ket{1 \dots 1}}$.
An even number of these cancel the $\bar{X}/\bar{Z}$ operation, while an odd number \revSelf{applies it}.
That is, 
\begin{equation}
    \begin{aligned}
        &\left( \prod_{k = 0}^i \overline{CO}_{\mathcal{S}_{i - k},\mathcal{Q}_1} \right) \ket{x}_{\revBoth{GSCX}_{a,b}} \ket{\psi}_{\mathcal{Q}_1} =\\
        &\revSelf{\frac{1}{\sqrt{2^{a-1}}}} \sum_{\substack{0 \leq \ell < 2^a \\ \mathrm{wt}(\ell) \equiv x \pmod 2}} \left( \bigotimes_{p = 0}^{a - 1} \ket{\ell_p}^{\otimes b} \right) \left( \left( \bar{O}_{\mathcal{Q}_1} \right)^{\mathrm{wts(\ell,i) \pmod 2}} \ket{\psi}_{\mathcal{Q}_1} \right),
    \end{aligned}
\end{equation}
where \revSelf{$\ell_p$ is the $p$th bit of $\ell$, $\mathrm{wt}(\ell)$ is the Hamming weight of the binary representation of $\ell$,} and $\mathrm{wts}(\ell,i) = \sum_{q=0}^{i} \left( \left\lfloor \frac{\ell}{2^q} \right\rfloor \bmod 2 \right)$ is the Hamming weight of the first $i+1$ digits of the binary representation of $\ell$.

Hence, a stabilizer $g_x(j)$, since it acts on \revSelf{adjacent} pairs of cat states, does not modify the subspaces unless $\revSelf{\mathrm{wts}(\ell,i) \pmod 2}$ changes, which only happens when $i = j$.
In that case, the \revSelf{affected basis states} will be switched, and we must add $\bar{O}_{\mathcal{Q}_1}$ to the stabilizer in order to account for this.

Because the modified stabilizer now acts on both the $\revBoth{GSCX}_{a,b}$ and $\mathcal{Q}_1$ codes, the codes may be considered as one large code that can independently correct $\lfloor \frac{d' - 1}{2} \rfloor$ errors in the $\revBoth{GSCX}_{a,b}$ partition and $\lfloor \frac{d - 1}{2} \rfloor$ errors in the $\mathcal{Q}_1$ partition, where $d$ is the distance of $\mathcal{Q}_1$ and $d' = \mathrm{min}(a,b)$ is the distance of $\revBoth{GSCX}_{a,b}$.
Take, for example, a $\mathcal{Q}_1$ code of $GSC_{3,3}$.
After performing a logical controlled-NOT gate controlled by the $\mathcal{S}_0$ subregister of the helper, the symplectic representation of the large code is shown in Tab.~\ref{tab:symplectic-large-code}.
As a syndrome example, the error $Z_0,X_9$ would produce a syndrome where only $g_6$ measures 1.
\begin{table*}
    \centering
    \caption{The symplectic representation of the modified stabilizer, $g_{12}$, after performing $\overline{CX}_{\mathcal{S}_0,\mathcal{Q}_1}$ during a $\overline{CX}_{\revBoth{GSCX}_{3,3},\mathcal{Q}_1}$ logical gate.
    Here, $\mathcal{Q}_1$ is encoded with $GSC_{3,3}$.
    $g_{j}$ represents the $j$th generator.
    Qubit numbers are ordered 0-17 from left to right in each X and Z section.}
    \label{tab:symplectic-large-code}
    \begin{tabular}{c||ccccccccc|ccccccccc}
        & \multicolumn{18}{c}{\textbf{X}} \\
        \hline
        & \multicolumn{9}{c|}{$\revBoth{GSCX}_{3,3}$} & \multicolumn{9}{c}{$\mathcal{Q}_1$} \\
        \hline
         $g_0$ & 0 & 0 & 0 & 0 & 0 & 0 & 0 & 0 & 0 
         & 0 & 0 & 0 & 0 & 0 & 0 & 0 & 0 & 0 \\
         $g_1$ & 0 & 0 & 0 & 0 & 0 & 0 & 0 & 0 & 0 
         & 0 & 0 & 0 & 0 & 0 & 0 & 0 & 0 & 0 \\
         $g_2$ & 0 & 0 & 0 & 0 & 0 & 0 & 0 & 0 & 0 
         & 0 & 0 & 0 & 0 & 0 & 0 & 0 & 0 & 0 \\
         $g_3$ & 0 & 0 & 0 & 0 & 0 & 0 & 0 & 0 & 0 
         & 0 & 0 & 0 & 0 & 0 & 0 & 0 & 0 & 0 \\
         $g_4$ & 0 & 0 & 0 & 0 & 0 & 0 & 0 & 0 & 0 
         & 0 & 0 & 0 & 0 & 0 & 0 & 0 & 0 & 0 \\
         $g_5$ & 0 & 0 & 0 & 0 & 0 & 0 & 0 & 0 & 0 
         & 0 & 0 & 0 & 0 & 0 & 0 & 0 & 0 & 0 \\
         $g_6$ & 0 & 0 & 0 & 0 & 0 & 0 & 0 & 0 & 0 
         & 0 & 0 & 0 & 0 & 0 & 0 & 0 & 0 & 0 \\
         $g_7$ & 0 & 0 & 0 & 0 & 0 & 0 & 0 & 0 & 0 
         & 0 & 0 & 0 & 0 & 0 & 0 & 0 & 0 & 0 \\
         $g_8$ & 0 & 0 & 0 & 0 & 0 & 0 & 0 & 0 & 0 
         & 0 & 0 & 0 & 0 & 0 & 0 & 0 & 0 & 0 \\
         $g_9$ & 0 & 0 & 0 & 0 & 0 & 0 & 0 & 0 & 0 
         & 0 & 0 & 0 & 0 & 0 & 0 & 0 & 0 & 0 \\
         $g_{10}$ & 0 & 0 & 0 & 0 & 0 & 0 & 0 & 0 & 0 
         & 0 & 0 & 0 & 0 & 0 & 0 & 0 & 0 & 0 \\
         $g_{11}$ & 0 & 0 & 0 & 0 & 0 & 0 & 0 & 0 & 0 
         & 0 & 0 & 0 & 0 & 0 & 0 & 0 & 0 & 0 \\
         $\mathbf{g_{12}}$ & \textbf{1} & \textbf{1} & \textbf{1} & \textbf{1} & \textbf{1} & \textbf{1} & 0 & 0 & 0 
         & 0 & 0 & 0 & 0 & 0 & 0 & 0 & 0 & 0 \\
         $g_{13}$ & 0 & 0 & 0 & 1 & 1 & 1 & 1 & 1 & 1 
         & 0 & 0 & 0 & 0 & 0 & 0 & 0 & 0 & 0 \\
         $g_{14}$ & 0 & 0 & 0 & 0 & 0 & 0 & 0 & 0 & 0 
         & 1 & 1 & 1 & 1 & 1 & 1 & 0 & 0 & 0 \\
         $g_{15}$ & 0 & 0 & 0 & 0 & 0 & 0 & 0 & 0 & 0 
         & 0 & 0 & 0 & 1 & 1 & 1 & 1 & 1 & 1 \\
         \hline
         \hline
         & \multicolumn{18}{c}{\textbf{Z}} \\
        \hline
        & \multicolumn{9}{c|}{$\revBoth{GSCX}_{3,3}$} & \multicolumn{9}{c}{$\mathcal{Q}_1$} \\
        \hline
         $g_0$ & 1 & 1 & 0 & 0 & 0 & 0 & 0 & 0 & 0 
         & 0 & 0 & 0 & 0 & 0 & 0 & 0 & 0 & 0 \\
         $g_1$ & 0 & 1 & 1 & 0 & 0 & 0 & 0 & 0 & 0 
         & 0 & 0 & 0 & 0 & 0 & 0 & 0 & 0 & 0 \\
         $g_2$ & 0 & 0 & 0 & 1 & 1 & 0 & 0 & 0 & 0 
         & 0 & 0 & 0 & 0 & 0 & 0 & 0 & 0 & 0 \\
         $g_3$ & 0 & 0 & 0 & 0 & 1 & 1 & 0 & 0 & 0 
         & 0 & 0 & 0 & 0 & 0 & 0 & 0 & 0 & 0 \\
         $g_4$ & 0 & 0 & 0 & 0 & 0 & 0 & 1 & 1 & 0 
         & 0 & 0 & 0 & 0 & 0 & 0 & 0 & 0 & 0 \\
         $g_5$ & 0 & 0 & 0 & 0 & 0 & 0 & 0 & 1 & 1 
         & 0 & 0 & 0 & 0 & 0 & 0 & 0 & 0 & 0 \\
         $g_6$ & 0 & 0 & 0 & 0 & 0 & 0 & 0 & 0 & 0 
         & 1 & 1 & 0 & 0 & 0 & 0 & 0 & 0 & 0 \\
         $g_7$ & 0 & 0 & 0 & 0 & 0 & 0 & 0 & 0 & 0 
         & 0 & 1 & 1 & 0 & 0 & 0 & 0 & 0 & 0 \\
         $g_8$ & 0 & 0 & 0 & 0 & 0 & 0 & 0 & 0 & 0 
         & 0 & 0 & 0 & 1 & 1 & 0 & 0 & 0 & 0 \\
         $g_9$ & 0 & 0 & 0 & 0 & 0 & 0 & 0 & 0 & 0 
         & 0 & 0 & 0 & 0 & 1 & 1 & 0 & 0 & 0 \\
         $g_{10}$ & 0 & 0 & 0 & 0 & 0 & 0 & 0 & 0 & 0 
         & 0 & 0 & 0 & 0 & 0 & 0 & 1 & 1 & 0 \\
         $g_{11}$ & 0 & 0 & 0 & 0 & 0 & 0 & 0 & 0 & 0 
         & 0 & 0 & 0 & 0 & 0 & 0 & 0 & 1 & 1 \\
         $\mathbf{g_{12}}$ & 0 & 0 & 0 & 0 & 0 & 0 & 0 & 0 & 0 
         & \textbf{1} & 0 & 0 & \textbf{1} & 0 & 0 & \textbf{1} & 0 & 0 \\
         $g_{13}$ & 0 & 0 & 0 & 0 & 0 & 0 & 0 & 0 & 0 
         & 0 & 0 & 0 & 0 & 0 & 0 & 0 & 0 & 0 \\
         $g_{14}$ & 0 & 0 & 0 & 0 & 0 & 0 & 0 & 0 & 0 
         & 0 & 0 & 0 & 0 & 0 & 0 & 0 & 0 & 0 \\
         $g_{15}$ & 0 & 0 & 0 & 0 & 0 & 0 & 0 & 0 & 0 
         & 0 & 0 & 0 & 0 & 0 & 0 & 0 & 0 & 0 \\
    \end{tabular}
\end{table*}

{\normalcolor \subsection{Number of Encoded Qubits}
First, we show that the combined code encodes the same total number of logical qubits as each code does separately.
$GSC_{a,b}$ is an $[[ab, 1, \min(a,b)]]$ code with $ab-1$ stabilizers.
Say that $\mathcal{Q}_1$ is encoded in an $[[n_{\mathcal{Q}_1}, k_{\mathcal{Q}_1}, d_{\mathcal{Q}_1}]]$ code, thereby having $g_{\mathcal{Q}_1} = n_{\mathcal{Q}_1} - k_{\mathcal{Q}_1}$ stabilizer generators~\cite{gottesmanStabilizers}.
The combined code keeps all the qubits and stabilizers from both codes, only modifying one stabilizer, thus having
\begin{equation}
    \begin{aligned}
        n_{\mathrm{combind}} &= n_{GSCX} + n_{\mathcal{Q}_1} \\
        &= ab + n_{\mathcal{Q}_1} \\
        g_{\mathrm{combined}} &= g_{GSCX} + g_{\mathcal{Q}_1} \\
        &= ab - 1 + n_{\mathcal{Q}_1} - k_{\mathcal{Q}_1}
    \end{aligned}
\end{equation}
qubits and stabilizers, respectively.
The generators remain independent because each modified $GSCX$ generator retains its original action on the $GSCX$ partition and acquires an additional factor only on the disjoint $\mathcal Q_1$ partition. 
Any dependence among the combined generators would therefore imply a dependence among the original $GSCX$ or $\mathcal{Q}_1$ stabilizer generators.
Hence, the combined code encodes
\begin{equation}
    \begin{aligned}
        k_{\mathrm{combined}} &= n_{\mathrm{combind}} - g_{\mathrm{combined}} \\
        &= ab + n_{\mathcal{Q}_1} - (ab - 1 + n_{\mathcal{Q}_1} - k_{\mathcal{Q}_1}) \\
        &= k_{\mathcal{Q}_1} + 1
    \end{aligned}
\end{equation}
qubits, preserving the total number of encoded qubits in both codes.

\subsection{Commutativity}
Here, we show the required property that all the stabilizers commute with each other holds in the combined code.
Firstly, all of the unmodified stabilizers from one code trivially commute with those from the other code since they act on separate qubits.
The only stabilizer we need to be concerned about is the modified stabilizer.

It performs only the unmodified stabilizer on the qubits it has in common with the $GSCX$ register, whereas the modification adds gates only on the other register.
Thus, it commutes with all unmodified $GSCX$ stabilizers.
Conversely, it performs the gates equal to a logical $\bar{O}$ operator on the qubits it has in common with the $\mathcal{Q}_1$ register.
Thus, since all $\mathcal{Q}_1$ stabilizers commute with $\bar{O}_{\mathcal{Q}_1}$ by definition, the modified stabilizer commutes with all $\mathcal{Q}_1$ stabilizers as well.

\subsection{Decoding and Partition Distance Preservation}\label{sec:decoding}
In this section, we verify that correctable errors in both partitions remain distinguishable.
The key insight is that, first, the stabilizers in $\mathcal{Q}_1$ remain unmodified, so the measurements of those stabilizers alone can continue to correct $t_{\mathcal{Q}_1}$ arbitrary errors in the $\mathcal{Q}_1$ partition.
The only stabilizer in $GSCX$ that needs additional consideration is the modified one.
It may anticommute with an error in either partition.
However, one can first determine the errors (or an error with an equivalent commutative effect) in the $\mathcal{Q}_1$ partition by the $\mathcal{Q}_1$ stabilizers alone and determine their effect on the modified stabilizer.
Then, one can recover the modified stabilizer's measurement as if it were unmodified and thus determine the errors in the $GSCX$ partition by including all of the unmodified $GSCX$ stabilizers.

More formally, we show that even if one partition is saturated with a correctable number of errors, the distance of the remaining partition remains as it was before being combined.
Say a decoder $\mathcal{D}_{\mathcal{Q}_1}: s_{\mathcal{Q}_1}  \rightarrow r_{\mathcal{Q}_1}$ takes an error syndrome $s_{\mathcal{Q}_1} \in \{1,-1\}^{g_{\mathcal{Q}_1}}$ consisting of $g_{\mathcal{Q}_1}$ stabilizer measurements and returns a recovery $r_{\mathcal{Q}_1} \in \{0,1\}^{2n_{\mathcal{Q}_1}}$ of $2n_{\mathcal{Q}_1}$ bits.
A 1 in the $i$th position in the first $n$ bits of the recovery represents a $Z$ error on the $i$th qubit of the $\mathcal{Q}_1$ partition.
A 1 in the $(n_{\mathcal{Q}_1} + i)$-th position represents an $X$ error on the $i$th qubit.
A 1 in both the $i$th and $(n_{\mathcal{Q}_1} + i)$-th position represents a $Y$ error, and a 0 represents no error.

Before any operations, each code acts independently and preserves its distance on its respective partition.
That is, the combined recovery $r_{c} = [r_{\mathcal{Q}_1}, r_{GSCX}]$ consists of the recovery $r_{\mathcal{Q}_1} = \mathcal{D}_{\mathcal{Q}_1}(s_{\mathcal{Q}_1})$ on the $\mathcal{Q}_1$ partition and $r_{GSCX} = \mathcal{D}_{GSCX}(s_{GSCX})$ on the $GSCX$ partition.

Say the combined code returns an error syndrome of $s_{c} = [s_{\mathcal{Q}_1}, s_{GSCX,\mathrm{unmod}}, s_{GSCX,\mathrm{mod}}]$, i.e., the measurement results from the $\mathcal{Q}_1$ stabilizers, the unmodified $GSCX$ stabilizers, and the modified $GSCX$ stabilizer, respectively.
None of the $\mathcal{Q}_1$ stabilizers are modified, so $\mathcal{D}_{\mathcal{Q}_1}(s_{\mathcal{Q}_1})$ can be run directly, preserving the distance of the $\mathcal{Q}_1$ partition, regardless of errors on the $GSCX$ partition.

The effect of $r_{\mathcal{Q}_1}$ on $s_{GSCX,\mathrm{mod}}$ can be efficiently determined by $(-1)^{r_{\mathcal{Q}_1} \cdot \bar{O}_{\mathcal{Q}_1}}$, where $\cdot$ is the vector inner product mod 2 and $\bar{O}_{\mathcal{Q}_1}$ is represented as a row in the symplectic matrix.
Thus, $s_{GSCX} = [s_{GSCX,\mathrm{unmod}}, (-1)^{r_{\mathcal{Q}_1} \cdot \bar{O}_{\mathcal{Q}_1}}s_{GSCX,\mathrm{mod}}]$ consists of the measurement results from the unmodified $GSCX$ stabilizers and the adjusted, modified $GSCX$ stabilizer.
We can thus run $\mathcal{D}_{GSCX}(s_{GSCX}) = \mathcal{D}_{GSCX}( [s_{GSCX,\mathrm{unmod}}, (-1)^{r_{\mathcal{Q}_1} \cdot \bar{O}_{\mathcal{Q}_1}}s_{GSCX,\mathrm{mod}}])$, preserving the distance of the $GSCX$ register.
Since we can run both decoders, $\mathcal{D}_{\mathcal{Q}_1}$ and $\mathcal{D}_{GSCX}$ as if they acted independently on their respective partitions, the distances in both partitions are preserved.

\subsection{Propagated Errors}\label{sec:propagated-errors}
During a transversal $\overline{CO}_{\mathcal{S}_i,\mathcal{Q}_1}$ gate, errors may propagate from one code to the other.
The possible propagation errors are listed in Table~\ref{tab:propagated-errors}, which were found with an exhaustive search using the method described in Ref.~\cite{nielsen_chuang_2010_unitary_gates_and_the_stabilizer_formalism}.
\begin{table}
    \centering
    \caption{Possible Error Propagation}
    \label{tab:propagated-errors}
    \begin{tabular}{Sl Sc Sc}
        \toprule
        \textbf{Gate} & \textbf{Initial Error} & \textbf{Propagated Error} \\
         & \textbf{Type / Register} & \textbf{Type / Register} \\
        \midrule
        $\overline{CX}$ & X / Control & X / Target \\
        $\overline{CZ}$ & X / Control & Z / Target \\
        $\overline{CX}$ & Z / Target & Z / Control \\
        $\overline{CZ}$ & X / Target & Z / Control \\
        \bottomrule
    \end{tabular}
\end{table}

While errors can propagate to both partitions, a single error in one partition can only propagate to at most a single location in the other partition.
Since the $\overline{CO}_{\mathcal{S}_i,\mathcal{Q}_1}$ gates happen in between QEC rounds, and we've seen that both the $GSCX$ and $\mathcal{Q}_1$ partitions are independently correctable and maintain their distances (see Sec.~\ref{sec:decoding}), fault-tolerance is preserved.
}

\section{Resource Overhead}\label{sec:overhead}

Consider (1) a maximum number of qubits, $n_{\mathcal{Q}_1}$, in a target code with distance $d_{\mathcal{Q}_1}$, (2) the number of qubits, $n_{RC}$, in a code with distance $d_{\mathcal{Q}_1}$ that has a fault-tolerant implementation of a logical $\bar{Z^\theta}$ gate, and (3) $n_{MC} = \max(n_{\mathcal{Q}_1}, n_{RC})$.
Then, a circuit using these codes to implement our SCG Clifford+$Z^\theta$ gate set incurs an overhead of up to
\begin{equation}\label{eq:bounds}
    \begin{aligned}
        m &= O(d_{\mathcal{Q}_1}) \\
        N^{(2)} &= O\left( \left( n_{MC} \right)^4 \right) \\
        N^{(1)} &= O\left( \left(d_{\mathcal{Q}_1}\right)^2 + n_{RC} \right) \\
        n &= O\left( (n_{MC})^3 \right) \\
    \end{aligned}
\end{equation}
QEC rounds, double-qubit physical gates, single-qubit physical gates, and qubits, respectively.
$m$, the additional number of QEC rounds, is due to the logical $\overline{CO}_{\revBoth{GSCX},\mathcal{Q}_1}$ gate being performed in $a$ sequential steps with QEC rounds performed after each one.
Since there are a constant number of $\overline{CO}_{\revBoth{GSCX},\mathcal{Q}_1}$ gates per SCG gate, and $a$ should generally be equal to $d_{\mathcal{Q}_1}$, $m$ scales as $O(d_{\mathcal{Q}_1})$.
The results of the other metrics are explained in the following sections.

\subsection{Assumptions and Definitions}

We define the following parameters, summarized in Table~\ref{tab:resource-overhead-parameters}, for the ancilla and target codes.
As in previous sections, $\revBoth{GSCX}_{a,b}$ represents $\revBoth{GSCX}$ with $a$ cat states and $b$ qubits per cat state.
$\revBoth{GSCX}_{a,b}$ therefore uses $n_{\revBoth{GSCX}} = ab$ data qubits and $s = ab - 1$ stabilizers.
Its distance is $d_{\revBoth{GSCX}} = \min(a, b)$, allowing $t_{\revBoth{GSCX}} = \lfloor (d_{\revBoth{GSCX}}-1)/2 \rfloor$ arbitrary errors to be corrected.
It follows, then, that $d_{\revBoth{GSCX}} \leq a, b$.

$\mathcal{Q}_1$, here representing the logical qubit encoded with the largest target QEC code used in the system, has $n_{\mathcal{Q}_1}$ data qubits and distance $d_{\mathcal{Q}_1}$.
$RC$, representing a code having a transversal implementation of a desired $\bar{Z}^{p}$ rotation, has $n_{RC}$ data qubits.
With no assumptions, the number of qubits per cat state $b$ must be enough to perform a logical $\bar{X}$ or $\bar{Z}$ gate on either $\mathcal{Q}_1$ or $RC$.
Thus, $b \leq n_{MC}$, where $n_{MC} = \max(n_{\mathcal{Q}_1}, n_{RC})$, since the number of gates in $\bar{X}$ or $\bar{Z}$ cannot exceed the number of qubits in the largest code.
To match the code distance so that $d_{\revBoth{GSCX}} = d_{RC} = d_{\mathcal{Q}_1}$, we set $a = d_{RC} = d_{\mathcal{Q}_1} \leq n_{\mathcal{Q}_1}$.
Thus, $a, b \leq n_{MC}$ and the correctable errors $t_{\revBoth{GSCX}} = t_{RC} = t_{\mathcal{Q}_1} < d_{\mathcal{Q}_1}$.

\begin{table}
    \centering
    \caption{Resource Overhead Parameters and Bounds}
    \label{tab:resource-overhead-parameters}
    \begin{tabular}{Sl Sc Sl}
        \toprule
        \textbf{Code} & \textbf{Data Qubits} & \textbf{Distance /} \\
         & & \textbf{Correctable Errors} \\
        \midrule
        $\mathcal{Q}_1$ & $n_{\mathcal{Q}_1}$ & $d_{\mathcal{Q}_1} \leq n_{\mathcal{Q}_1}$ \\
        & & $t_{\mathcal{Q}_1} < d_{\mathcal{Q}_1}$ \\
        \hline
        $RC$ & $n_{RC}$ & $d_{RC} = d_{\mathcal{Q}_1}$ \\
        & & $t_{RC} < d_{\mathcal{Q}_1}$ \\
        \hline
        $TC$ & $n_{TC}$ & $d_{TC} = d_{\mathcal{Q}_1}$ \\
        & $= O\left((d_{\mathcal{Q}_1})^2 \log^2 d_{\mathcal{Q}_1}\right)$ & $t_{TC} < d_{\mathcal{Q}_1}$ \\
        \hline
        $MC$ & $n_{MC} = \max(n_{\mathcal{Q}_1}, n_{RC})$ & $d_{MC} = d_{\mathcal{Q}_1}$ \\
        & & $t_{MC} < d_{\mathcal{Q}_1}$ \\
        \hline
        $\revBoth{GSCX}_{a,b}$ & $n_{\revBoth{GSCX}} = ab$ & $d_{\revBoth{GSCX}} = d_{\mathcal{Q}_1}$ \\
        & $\leq (n_{MC})^2$ & $t_{\revBoth{GSCX}} < d_{\mathcal{Q}_1}$ \\
        \bottomrule
    \end{tabular}
\end{table}

The overhead associated with initial encoding, initial state preparation, and final X-basis measurements are excluded, as these are implementation-specific. 
Error correction overhead for the target codes (encoding $\mathcal{Q}_1, \mathcal{Q}_2, \dots$) are also excluded, as these are considered part of the underlying system with or without SCG gates.
We assume a Shor-style stabilizer measurement procedure, requiring 
\begin{itemize}
    \item $r = d_{MC} \leq n_{MC}$ measurement rounds per QEC round~\cite{numMeasurementRoundsForShorStyle}.
    \item One double-qubit gate for every gate in the stabilizer being measured.
    \item Up to $t_{MC} = \lfloor (d_{MC} - 1) / 2 \rfloor$ single-qubit gates per QEC round for error recovery.
    \item One ancilla qubit for every non-identity gate in the stabilizer being measured. 
\end{itemize}

\subsection{Double-Qubit Gate Overhead}
Out of the $s$ stabilizers of $\revBoth{GSCX}$, $z = a(b - 1)$ of them have $2$ gates and $x = a - 1$ of them have $2b$ gates. 
Assuming the syndrome measurement procedure requires a double-qubit controlled physical gate per stabilizer gate, the total number of double-qubit gates for syndrome extraction is 
\begin{equation}
    \begin{aligned}
        N^{(2)}_{\revBoth{GSCX}} &= r(2z + 2bx) \\
        &= r(2a(b - 1) + 2b(a - 1)) \\
        &= O(rab) \\
        &\leq O\left( (n_{MC})^3 \right).
    \end{aligned}
\end{equation}
Similarly, a QEC round for $RC$ incurs an overhead of $N^{(2)}_{RC} \leq \left(n_{RC}\right)^3$ double-qubit physical gates because there are up to $n_{RC}$ stabilizers, each with up to $n_{RC}$ gates, and $r \leq n_{RC}$ measurement rounds per QEC round.

Logical controlled-$\bar{X}/\bar{Z}$ gates controlled by $\revBoth{GSCX}$ ($\overline{CO}_{\revBoth{GSCX}, MC}$) are executed in $a = d_{MC} \leq n_{MC}$ iterations.
Each iteration performs up to $b \leq n_{MC}$ double-qubit physical gates followed by error correction with a modified stabilizer.
The modified stabilizer adds $|\bar{O}_{MC}| \leq n_{MC}$ double-qubit physical gates to that of the unmodified syndrome extraction, $N^{(2)}_{\revBoth{GSCX}}$.
Thus, the total double-qubit physical gate overhead for $\overline{CO}_{\revBoth{GSCX}, MC}$ after $a$ iterations is:
\begin{equation}
    \begin{aligned}
        N^{(2)}_{\overline{CO}_{\revBoth{GSCX},MC}} &= a \left( b + N^{(2)}_{\revBoth{GSCX}} + r|\bar{O}_{MC}| \right) \\
        &\leq n_{MC} \left( n_{MC} + (n_{MC})^3 + \left( n_{MC} \right)^2 \right) \\
        &= O\left( (n_{MC})^4 \right).
    \end{aligned}
\end{equation}

The SCG Hadamard gate performs a logical $\overline{CX}_{\revBoth{GSCX},MC}$ gate followed by a logical $\overline{CZ}_{\revBoth{GSCX},MC}$ gate, resulting in $N^{(2)}_{\bar{H}} = 2 N^{(2)}_{\overline{CO}_{\revBoth{GSCX},MC}} = O\left((n_{MC})^4\right)$ double-qubit physical gates.
The SCG controlled-$\bar{X}/\bar{Z}$ gates perform a logical $\overline{CZ}_{\revBoth{GSCX},\mathcal{Q}_1}$ gate, then an SCG $\bar{H}_{\revBoth{GSCX}}$ gate, then a logical $\overline{CO}_{\revBoth{GSCX},\mathcal{Q}_2}$ gate.
Combined,
\begin{equation}
    \begin{aligned}
        N^{(2)}_{\overline{CO}_{\mathcal{Q}_1,\mathcal{Q}_2}} &= N^{(2)}_{\overline{CZ}_{\revBoth{GSCX},\mathcal{Q}_1}} + N^{(2)}_{\bar{H}} + N^{(2)}_{\overline{CO}_{\revBoth{GSCX},\mathcal{Q}_2}} \\
        &= O\left((n_{\mathcal{Q}_1})^4\right)
    \end{aligned}
\end{equation}
double-qubit physical gates are required.

\revSelf{The SCG Z-rotation gate variation that utilizes the SCG controlled-$\bar{X}$ performs an SCG $\overline{CX}_{\mathcal{Q}_1,RC}$ gate, while the alternative utilizes an SCG $\bar{H}_{RC}$ gate followed by a measurement with error correction.
Both variations are completed with a logical $\left(\bar{Z}^{p}\right)_{RC}$ gate with error correction.}
The logical $\left(\bar{Z}^{p}\right)_{RC}$ gate only uses single-qubit physical gates and therefore does not contribute to the double-qubit physical gate overhead.
Thus, the SCG Z-rotation gate has a double-qubit physical gate overhead of 
\begin{equation}
    \begin{aligned}
        N^{(2)}_{\bar{Z}^{p}} &= \begin{cases} 
        N^{(2)}_{\overline{CX}_{\mathcal{Q}_1,RC}} + N^{(2)}_{RC} & \text{Variation 1} \\
        N^{(2)}_{\bar{H}_{RC}} + 2 N^{(2)}_{RC}  & \text{Variation 2} \end{cases}  \\
        &= O\left((n_{RC})^4 + \left(n_{RC}\right)^3\right) \\
        &= O\left( (n_{RC})^4 \right).
    \end{aligned}
\end{equation}
The double-qubit physical gate overhead is summarized in Table~\ref{tab:double-qubit-gate-overhead}.
\begin{table}
    \centering
    \caption{Double-qubit gate overhead.}
    \label{tab:double-qubit-gate-overhead}
    \begin{tabular}{Sc Sc}
        \toprule
        \textbf{Gate Overhead} & \textbf{Gate Bound} \\
        \midrule
        $N^{(2)}_{\revBoth{GSCX}}$ & $O\left( (n_{MC})^3 \right)$ \\
        \hline
        $N^{(2)}_{RC}$ & $O\left( \left(n_{RC}\right)^3 \right)$ \\
        \hline
        $N^{(2)}_{\overline{CO}_{\revBoth{GSCX},MC}}$ & $O\left( (n_{MC})^4 \right)$ \\
        \hline
        $N^{(2)}_{\bar{H}}$ & $O\left((n_{MC})^4\right)$ \\
        \hline
        $N^{(2)}_{\overline{CO}_{\mathcal{Q}_1,\mathcal{Q}_2}}$ & $O\left((n_{\mathcal{Q}_1})^4\right)$ \\
        \hline
        $N^{(2)}_{\bar{Z}^{p}}$ & $O\left( (n_{RC})^4 \right)$ \\
        \bottomrule
    \end{tabular}
\end{table}

\subsection{Single-Qubit Gate Overhead}
Error recovery requires single-qubit physical gates (controlled by classical results), generally applying one single-qubit physical gate per error.
Hence, error correction for $\revBoth{GSCX}$ requires up to $N^{(1)}_{\revBoth{GSCX}} = t_{\revBoth{GSCX}} < d_{\mathcal{Q}_1}$ single-qubit physical gates.
Similarly, $RC$ requires $N^{(1)}_{RC} = t_{RC} < d_{\mathcal{Q}_1}$ single-qubit physical gates.

The rest of the single-qubit physical gate overhead follows the same logic as the double-qubit gate overhead, except $\left(\bar{Z}^{p}\right)_{RC}$ additionally uses up to $n_{RC}$ single-qubit physical gates for the additional $RC$ register.
The single-qubit physical gate overhead is summarized in Table~\ref{tab:single-qubit-gate-overhead}.
\begin{table}
    \centering
    \caption{Single-qubit gate overhead.}
    \label{tab:single-qubit-gate-overhead}
    \begin{tabular}{Sc Sc}
        \toprule
        \textbf{Gate Overhead} & \textbf{Expression /} \\
         & \textbf{Gate Bound} \\
        \midrule
        $N^{(1)}_{\revBoth{GSCX}}$ & $t_{\revBoth{GSCX}}$ \\
         & $O(d_{\mathcal{Q}_1})$ \\
        \hline
        $N^{(1)}_{RC}$ & $t_{RC}$ \\
         & $O(d_{\mathcal{Q}_1})$ \\
        \hline
        $N^{(1)}_{\overline{CO}_{\revBoth{GSCX},MC}}$ & $a N^{(1)}_{\revBoth{GSCX}}$ \\
         & $O \left((d_{\mathcal{Q}_1})^2 \right)$ \\
        \hline
        $N^{(1)}_{\bar{H}}$ & $2 N^{(1)}_{\overline{CO}_{\revBoth{GSCX},MC}}$ \\
         & $O\left((d_{\mathcal{Q}_1})^2\right)$ \\
        \hline
        $N^{(1)}_{\overline{CO}_{\mathcal{Q}_1,\mathcal{Q}_2}}$ & $N^{(1)}_{\overline{CZ}_{\revBoth{GSCX},\mathcal{Q}_1}} + N^{(1)}_{\bar{H}} + N^{(1)}_{\overline{CO}_{\revBoth{GSCX},\mathcal{Q}_2}}$ \\
         & $O\left((d_{\mathcal{Q}_1})^2\right)$ \\
        \hline
        $N^{(1)}_{\bar{Z}^{p}}$ & $N^{(1)}_{\bar{H}_{RC}} + n_{RC} + 2 N^{(1)}_{RC}$ \\
         & $O\left( (d_{\mathcal{Q}_1})^2 + n_{RC} \right)$ \\
        \bottomrule
    \end{tabular}
\end{table}

\subsection{Qubit Overhead}
The qubit overhead for each ancilla code consists of $n_C \leq n_{d} + n_{a} n_{s}$, where $n_d$ is the number of data qubits for that code, $n_a$ is the number of ancilla qubits required to measure the largest stabilizer, and $n_s$ is the number of stabilizers in the code.
This is a conservative estimate, assuming all ancilla qubits exist simultaneously, as if the stabilizer measurements were all done in parallel.
The largest of the $\revBoth{GSCX}$ stabilizers is the modified X-stabilizer during a logical $\overline{CO}_{\revBoth{GSCX},MC}$ gate, which has $2b + |\bar{O}_{MC}| = O(n_{MC})$ gates.
Then, as a Shor-style measurement requires one qubit per gate, the total number of additional qubits required per $\revBoth{GSCX}$ ancilla register is up to
\begin{equation}
    \begin{aligned}
        n_{\overline{CO}_{\revBoth{GSCX},MC}} &= O(n_{\revBoth{GSCX}} + n_{MC} (ab - 1)) \\
        &= O((n_{MC})^2 + (n_{MC})^3) \\
        &= O\left( (n_{MC})^3 \right).
    \end{aligned}
\end{equation}
Furthermore, each $RC$ register cannot have more than $n_{RC}$ stabilizers and $n_{RC}$ gates per stabilizer, thus requiring $n_{\left( \bar{Z}^{p} \right)_{RC}} = O\left( (n_{RC})^2 \right)$ total additional qubits.

Table~\ref{tab:qubit-overhead} lists the qubit overhead associated with each SCG gate and the components contributing to it.
\begin{table}
    \centering
    \caption{Qubit overhead for SCG gates.}
    \label{tab:qubit-overhead}
    \begin{tabular}{Sc Sc Sc Sc Sc}
        \toprule
        \textbf{Gate} & \textbf{\# of \revBoth{GSCX}} & \textbf{\# of RC} & \textbf{Expression /} \\
         & & & \textbf{Qubit Bound} \\
        \midrule
        $\bar{H}$ & 1 & 0 & $n_{\overline{CO}_{\revBoth{GSCX},MC}}$ \\
         & & & $O\left( (n_{MC})^3 \right)$ \\
        \hline
        $\overline{CO}_{\mathcal{Q}_1,\mathcal{Q}_2}$ & 2 & 0 & $2 n_{\overline{CO}_{\revBoth{GSCX},\mathcal{Q}_1}}$ \\
         & & & $O\left( (n_{\mathcal{Q}_1})^3 \right)$ \\
        \hline
        $\bar{Z}^{p}$ & 2 & 1 & $2 n_{\overline{CO}_{\revBoth{GSCX},MC}} +  n_{\left( \bar{Z}^{p} \right)_{RC}}$ \\
        (Variation 1) & & & $O\left( (n_{MC})^3 \right)$ \\
        \hline
        $\bar{Z}^{p}$ & 1 & 1 & $n_{\overline{CO}_{\revBoth{GSCX},MC}} + n_{\left( \bar{Z}^{p} \right)_{RC}}$ \\
        (Variation 2) & & & $O\left( (n_{MC})^3 \right)$ \\
        \bottomrule
    \end{tabular}
\end{table}

\subsection{Triorthogonal Bounds}\label{sec:triorthogonal-bounds}

To determine the bounds given in Eq.~\eqref{eq:bounds} for the Clifford+$T$ gate set, $RC = TC$, one can use a triorthogonal code.
Based on known constructions for triorthogonal codes~\cite{triorthogonal}, the number of qubits $n_{TC}$ required for a target distance $d_{TC} = d_{\mathcal{Q}_1}$ is bounded by
\begin{equation}
    \begin{aligned}
        n_{TC} &= O\left(\left(d_{\mathcal{Q}_1}\right)^2 \log^2 d_{\mathcal{Q}_1}\right).
    \end{aligned}
\end{equation}

Haah and Hastings presented efficient protocols to generate triorthogonal matrices with distance $d_{T} = \Omega  \left( \frac{\sqrt{n_{T}}}{\log n_{T}} \right)$~\cite{triorthogonal_generation}.
This implies that
\begin{align}
    d_{T} &= \Omega  \left( \frac{\sqrt{n_{T}}}{\log n_{T}} \right) \\
    \frac{\sqrt{n_{T}}}{\log n_{T}} &= O\left( d_{T} \right) \\
    \sqrt{n_{T}} &= O\left( d_{T} \log n_{T} \right) \\
    n_{T} &= O\left( \left( d_{T} \right)^2 \log^2 n_{T} \right) \\
\end{align}
To get $\log n_{T}$ in terms of $d_{T}$,
\begin{align}
    n_{T} &= O\left( \left( d_{T} \right)^2 \log^2 n_{T} \right) \\
    \log n_{T} &= O\left( \log\left(\left( d_{T} \right)^2\right) + \log \left( \log^2 n_{T} \right) \right) \\
    &= O\left( 2\log d_{T} + 2 \log \left( \log n_{T} \right) \right) \\
    &= O\left( \log d_{T} \right) \\
\end{align}
Thus,
\begin{align}
    n_{T} &= O\left( \left( d_{T} \right)^2 \log^2 n_{T} \right) \\
    &= O\left( \left( d_{T} \right)^2 \log^2 d_{T} \right)
\end{align}

\end{document}